\documentclass[article]{IEEEtran}
\usepackage[utf8]{inputenc}
\usepackage{cite}
\usepackage{graphicx}
\usepackage{caption}
\usepackage{enumitem}
\PassOptionsToPackage{hyphens}{url}\usepackage{hyperref}
\usepackage{color, colortbl}
\definecolor{Gray}{gray}{0.9}
\usepackage{array}
\usepackage{soul}
\usepackage{float}
\usepackage{stfloats}

\newcolumntype{P}[1]{>{\centering\arraybackslash}p{#1}}
\title{A Survey on Security Attacks and Defense Techniques for Connected and Autonomous Vehicles}

\author{
    \IEEEauthorblockN{Minh Pham\IEEEauthorrefmark{1},
    Kaiqi Xiong\IEEEauthorrefmark{2}\\}
    \IEEEauthorblockA{
    \IEEEauthorrefmark{1}Department of Computer Science and Engineering, University of South Florida\\
    \IEEEauthorrefmark{2}Intelligent Computer Networking and Security Laboratory, 
    University of South Florida\\
    Email: \IEEEauthorrefmark{1}minhpham@mail.usf.edu,
    \IEEEauthorrefmark{2}xiongk@usf.edu
    }
}

\begin{document}
\maketitle
\begin{abstract}
    Autonomous Vehicle has been transforming intelligent transportation systems. As  telecommunication  technology  improves, autonomous vehicles are getting connected to each other and to infrastructures, forming Connected and Autonomous Vehicles (CAVs). CAVs will help humans achieve safe, efficient, and autonomous transportation systems. However, CAVs will face significant security challenges because many of their components are vulnerable to attacks, and a successful attack on a CAV may have significant impacts on other CAVs and infrastructures due to their communications. In this paper, we conduct a survey on 184 papers from 2000 to 2020 to understand state-of-the-art CAV attacks and defense techniques. This survey first presents a comprehensive overview of security attacks and their corresponding countermeasures on CAVs. We then discuss the details of attack models based on the targeted CAV components of attacks, access requirements, and attack motives. Finally, we identify some current research challenges and trends from the perspectives of both academic research and industrial development. Based on our studies of academic literature and industrial publications, we have not found any strong connection between academic research and industry's implementation on CAV-related security issues. While efforts from CAV manufacturers to secure CAVs have been reported, there is no evidence to show that CAVs on the market have the ability to defend against some novel attack models that the research community has recently found. This survey may give researchers and engineers a better understanding of the current status and trend of CAV security for CAV future improvement.
    
\end{abstract}

\section{Introduction}

Autonomous Vehicle (AV) has been a fascinating and impactful application of modern technology and it has been transforming human's intelligent transportation systems\cite{figueiredo2001towards,becker2000sensor}. 
As telecommunication technology improves, the concept of Connected Vehicles (CVs), which is the idea to connect vehicles and to communicate with road infrastructures and the Internet, has been realized and often implemented together with Autonomous vehicle \cite{bansal2017forecasting,uhlemann2015autonomous}. Many research studies in academia and industry have advanced Connected and Autonomous Vehicles (CAVs), aiming toward a safe, driverless, and efficient transportation system. These advancements have led to prominent public demonstrations of CAVs in North America, Japan, and Europe \cite{williams1988prometheus,benmimoun2009demonstration,suzuki2010development,van2012cooperative}. There are various levels of CAV automation, ranging from non-automated to fully automated. In 2018, the  Society of Automobile Engineers updated the official reference that specifically described the five levels of vehicle automation  \cite{sae2018taxonomy}. These five levels include Level 0-no automation, Level 1-driver assistance, Level 2-partial automation, Level 3-conditional automation, Level 4-high automation, and Level 5-full automation. In this paper, we consider automation levels 2, 3, 4, and 5, which are described by the official document \cite{sae2018taxonomy} that humans are not fully involved when the automated driving features are engaged (such as being hands-off).

A CAV consists of many sensing components, such as laser, radar, camera, Global Positioning System (GPS), and light detection and ranging (LiDAR) \cite{wyglinski2013security}, as well as their connection mechanisms, such as cellular connection, Bluetooth, IEEE 802.11p Wireless Access in Vehicular Environments (WAVE) \cite{ieee2010ieee}, and Wi-Fi. The sensing components enable a CAV to navigate in an environment with unknown obstacles. Using data from these sensing sensors, the surrounding environment and the vehicle’s location are computed by a system in a process known as Simultaneous Localization and Mapping (SLAM) \cite{durrant2006simultaneous,bailey2006simultaneous}. Connection mechanisms improve the driving experience or enhance an autonomous driving system by providing advanced knowledge and a bigger picture of the environment. Applications that utilize connection mechanisms include Intelligent Driver-Assistance Systems (IDAS) \cite{pilipovic2014toward}, safety features through Vehicle-to-Infrastructure communications \cite{bohm2008supporting}, and safety features through Vehicle-to-Vehicle communications \cite{xu2004vehicle,biswas2006vehicle}. While the sensing components and connection mechanisms have offered significant improvements in safety, cost, and fuel efficiency, they also created more opportunities for cyberattacks. 

\begin{table*}[bp]
	\caption{A comparison of survey papers on CAVs}
	\begin{center}
		
		\begin{tabular}{|>{\raggedright\arraybackslash}P{2cm} | P{2cm}| P{2cm} | P{2cm} | P{2cm} | P{2cm} | P{2cm} |}
			\hline
			\multicolumn{1}{|>{\centering\arraybackslash}m{2cm}|}{\cellcolor{Gray}\textbf{Survey Paper}}
			& \multicolumn{1}{>{\centering\arraybackslash}m{2cm}|}{\cellcolor{Gray} \textbf{Year Published}}
			& \multicolumn{1}{>{\centering\arraybackslash}m{2cm}|}{\cellcolor{Gray} \textbf{Reference Count}}
			& \multicolumn{1}{>{\centering\arraybackslash}m{2cm}|}{\cellcolor{Gray} \textbf{Year of Latest Reference}}
			& \multicolumn{1}{>{\centering\arraybackslash}m{2cm}|}{\cellcolor{Gray} \textbf{Focused Topic}}    
			& \multicolumn{1}{>{\centering\arraybackslash}m{2cm}|}{\cellcolor{Gray} \textbf{High-level Taxonomy}}
			& \multicolumn{1}{>{\centering\arraybackslash}m{2cm}|}{\cellcolor{Gray} \textbf{Outlining Open-issues}}\\
			\hline

			\centering Miller and Valasek \cite{miller2014survey} & 2014 & 12 & 2012 & CAVs & No & No\\ 
			\hline
			\centering Thing and Wu \cite{thing2016autonomous} & 2016 & 16 & 2016 & CAVs & Yes & No \\
			\hline
			\centering Haider et al. \cite{haider2016survey} & 2016 & 10 & 2015 & Global Positioning System & No & No \\
			\hline
			\centering Parkinson et al. \cite{parkinson2017cyber} & 2017 & 91 & 2016 & CAVs & No & Yes \\
			\hline
			\centering Tomlinson et al. \cite{tomlinson2018towards} & 2018 & 43 & 2018 & Controller Area Network & Yes & Yes\\
			\hline
			\centering van der Heijden et al. \cite{van2018survey} & 2018 & 126 & 2018 & Misbehavior detection in communication between CAVs & No & Yes\\
			\hline
			\centering This paper & 2020 & 184 & 2020 & CAVs & Yes & Yes \\
			\hline
		\end{tabular}
	\end{center}
\end{table*}

Attempts to deploy and test CAVs have been carried out in many places, and they are supported by governments and corporations. In September 2016, the United States Department of Transportation started the Connected Vehicle Pilot Deployment Program \cite{gopalakrishna2015connected,kitchener2017connected}, providing over 45 million USD to Wyoming \cite{wyoming}, New York City \cite{NYC}, and Tampa \cite{Tampa} to begin building connected vehicle programs. In the United Kingdom, the Centre for Connected and Autonomous Vehicles has invested 120 million GBP to support over 70 CAVs projects, with a further 68 million GBP coming from industry contributions \cite{UK}. In China, industry officials estimated that by 2035, there will be around 8.6 million autonomous vehicles on the road, of which 5.2 million are semi-autonomous (SAE levels 3 and 4) and 3.4 million are fully autonomous (SAE level 5) \cite{west2016moving}. In Japan, prime minister Shinzo Abe claimed to grow a fleet of thousands of autonomous vehicles to serve in Tokyo Olympics 2020 \cite{west2016moving}. In South Korea, two competitions were sponsored by Hyundai Motor Group to stimulate the development of CAVs \cite{jo2015development}. These two competitions were held in 2010 and 2012, respectively. The development of CAVs is gaining significant public attention. The unfortunate side effect of this public attention is that CAVs will probably become attractive targets for cyberattacks.

Furthermore, CAV engineers and manufacturers need to have a systematic understanding of the cybersecurity implications of CAVs. Even though no significant cyberattack has occurred to the publicly deployed CAV programs, there are potential security threats to CAVs that have been discovered largely by the academic research community \cite{petit2014potential}. These potential security attacks will be more harmful than attacks on non-automated transportation systems because drivers may not be mentally or physically available to take over the driving, and engineers and technicians may not be available immediately to recover a compromised system.

Considerable research efforts have been carried out for identifying vulnerabilities in CAVs, recommending potential mitigation techniques, and highlighting the potential impacts of cyberattacks on CAVs and related infrastructures \cite{checkoway2011comprehensive,koscher2010experimental,yadav2016security,petit2015remote}. Researchers have identified many vulnerabilities associated with sensors, electronic control units, and connection mechanisms. Some even demonstrated successful cyberattacks on CAVs and their components that are currently being sold and operated \cite{yan2016can,cao2019adversarial,Tesla3Spoof}. Since detailed and security-focused studies for CAVs are fairly new in the literature (the majority of technical and in-depth papers discussed in this survey are published after 2011), there is an absence of a comprehensive survey paper that utilizes the current literature to build a taxonomy and to suggest significant gaps and challenges. For example, Miller and Valasek (2014) \cite{miller2014survey} published a survey paper on attack surfaces but did not have much cover on defense strategies. Thing and Wu (2016) proposed a taxonomy of attacks and defenses but failed to point out specific examples in the literature with only 16 references \cite{thing2016autonomous}. Other survey papers that are dedicated to specific components of CAVs \cite{tomlinson2018towards,haider2016survey,van2018survey}. To our knowledge, Parkinson et al. (2017) \cite{parkinson2017cyber} is a state-of-the-art survey paper on this topic. Parkinson et al.'s paper reviewed 89 publicly accessible publications and identified knowledge gaps in the literature. However, we found that the authors missed interesting and important papers on some attack models and defense strategies, such as ones that we will cover in the GPS spoofing attacks \cite{yang2016gnss,o2013real} (section III-F), defense against LiDAR spoofing \cite{wang2015pseudorandom,nouri2017target} (section III-D), and adversarial input attack on cameras \cite{sitawarin2018darts,man2019poster} (section III-G). Meanwhile,  Parkinson et al. \cite{parkinson2017cyber} did not include any literature published after 2017. We tried our best to explore and present such technical papers, which are experimented not only on CAVs but also on related cyber-physical systems (e.g., unmanned aerial vehicles). Besides, our survey paper covers the recent developments of attacks and defenses on CAVs, including three ethical hacking studies on Tesla and Baidu autonomous vehicles in 2019. A comparison of survey papers can be found in Table I.

This survey paper aims to review published papers and technical reports on cybersecurity vulnerabilities and defenses of CAVs, to provide readers with a summary of past research efforts, to organize them into systematic groups, and to identify research gaps and challenges. We have surveyed 184 papers from 2000 to 2020 about CAVs and CAV components to understand the security challenges of CAVs. The first paper related to the security of CAVs was published in 2005 regarding secure software update for CAVs \cite{mahmud2005secure}. Since then, we observed an increasing trend of publications on this issue. From 2015 to 2019, we counted 8, 11, 14, 17, and 12 published papers per respective year about CAV-related security issues. We hope that our work can inform current and aspiring researchers and engineers of the security issues of CAVs as well as state-of-the-art defense and mitigation techniques. We further hope that our work can motivate other researchers to address cybersecurity challenges facing the development of CAVs. We acknowledge that research in this area is growing at a rapid rate. We also realize that some achievements from academia and industry might have been overlooked or not yet published. As a result, we observe that many vulnerabilities do not have enough tested solutions. Given the vast investment and rapid changes in the CAV industry, many individuals and corporations may not agree with our observations in this article, but any debate and criticism would be welcomed and appreciated for the growth of the community.

The remainder of this paper is organized as follows. Section 2 describes the taxonomies of attack and defense according to the components of CAVs, where those components are also explained in details. This section would provide readers a brief overview of the attack and defense techniques appeared in the existing literature so that readers without technical experience can have a high-level understanding of those attacks and defenses.  Section 3 discusses the attack techniques, their corresponding mitigation/defense techniques, and the challenges of defenses and the gaps between attack models and defense techniques. In section 4, we identify the trends, challenges, and open issues in academic research and industry developments. 

\section{Taxonomy of Attacks and Defenses}
The purpose of this section is to provide a high-level overview of the types of attacks and defenses that have been discussed for CAVs. In this section, we attempt to classify attack models and defense strategies based on their characteristics, but do not provide technical details. Instead, technical details of attack models and defense strategies with their corresponding references are presented in section III. Figures 3 and 4, which are presented in section II to point readers to related parts in section III, may help readers navigate easily between the two sections.

\subsection{Taxonomy of Attacks}
In this section, we discuss the CAV components whose vulnerabilities have been found in the literature and provide a high-level overview of the attack models.
\subsubsection{Attack targets}
As discussed in section I, a good CAV system consists of many sensor components and connection mechanisms. They work together and contribute to CAVs' functioning. Compromising or tampering any of these components may destabilize a CAV and serve the attacker's goal, such as stealing information and causing property damage and bodily injury. In this subsection, we describe CAV components that have been targeted by cyber attackers in the literature. Some attack models have been demonstrated as realistic threats while the others have only been discussed theoretically. While section II presents the classification of attacks and defense, section III gives the detailed discussion of attack models and defense strategies with their citations. Figure 1 summarizes the attack targets along with their corresponding subsections in section III, where readers can find the detailed discussions of the attack models and the mitigation techniques along with their references.

\begin{figure*}
  \centering
  \captionsetup{justification=centering}
  \includegraphics[width=\textwidth]{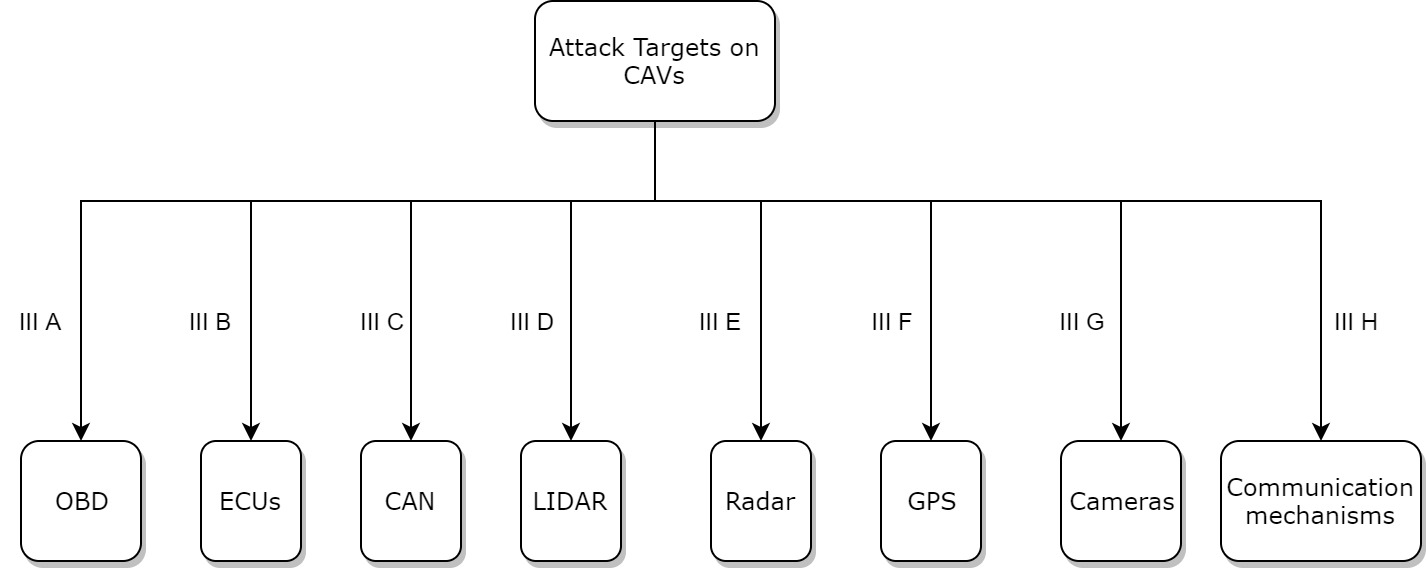}
  \caption{Possible attack targets on CAVs}
\end{figure*}

\textbf{On-board Diagnostic Port (OBD)} is a connection port that anyone can use to collect information about a vehicle's emissions, mileage, speed, and data on a vehicle's components. There are two OBD standards, namely OBD-I and OBD-II. OBD-I was introduced in 1987 but had many flaws, so it was replaced by OBD-II introduced in 1996 \cite{kalmeshwar2017development}. OBD-II port should be found in almost any modern vehicle, and CAVs are not exceptions. Figure 2 shows the OBD port on a Tesla Model X (SAE level 2). Modern OBD ports can provide real-time data \cite{lin2005development}. OBD also provides a pathway to acquire data from CAV's Electronic Control Units and possibly to modify the software embedded in those control units. Many manufacturers also use OBD ports to perform firmware updates \cite{checkoway2011comprehensive}. Attack models on OBD ports and their corresponding mitigation techniques are discussed in Section III-A.

\begin{figure}
  \centering
  \captionsetup{justification=centering}
  \includegraphics[width=0.5\textwidth]{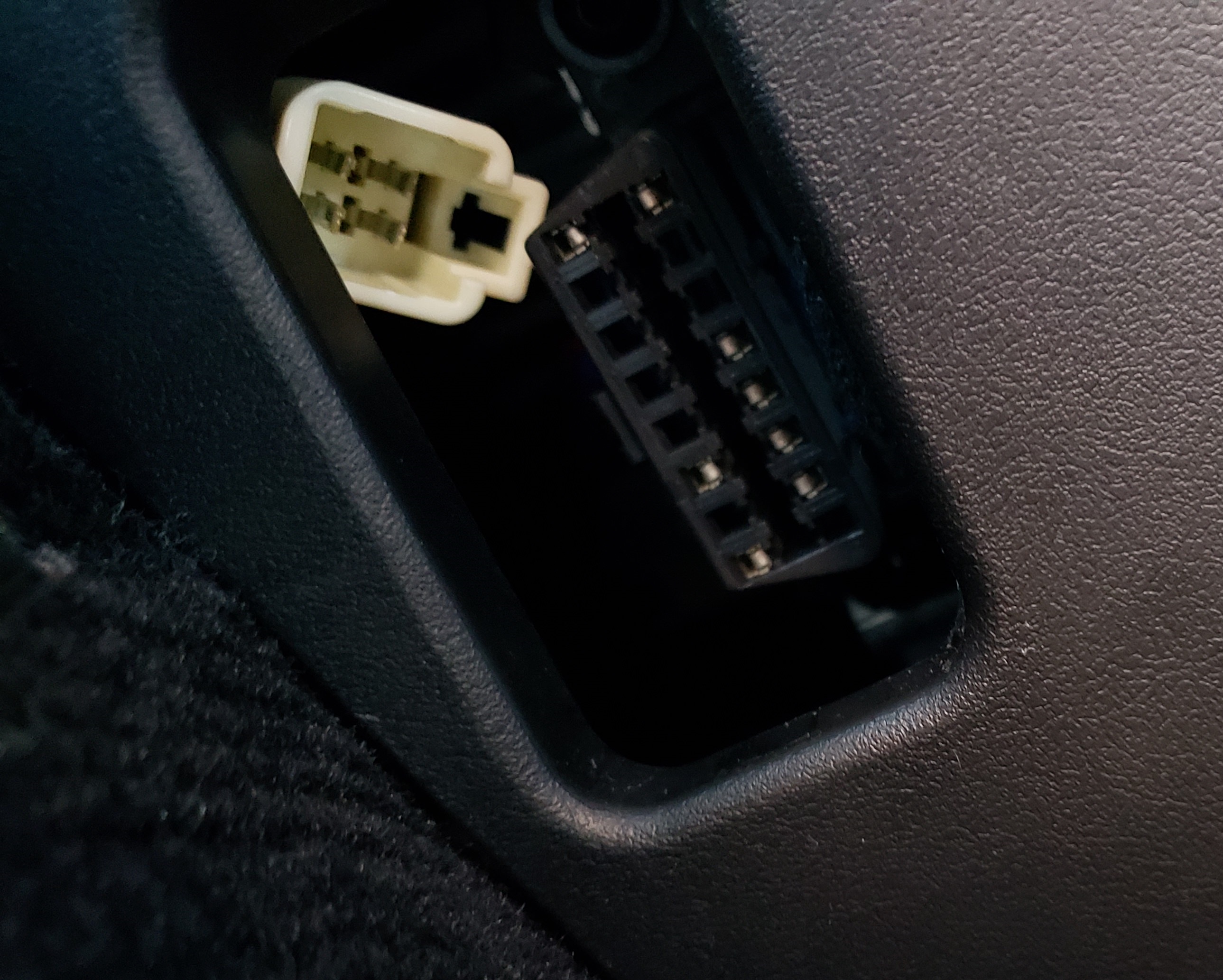}
  \caption{OBD port on Tesla Model X}
\end{figure}

\textbf{Electronic Control Units (ECUs)} are embedded electronic systems that control other subsystems in a vehicle. All modern vehicles use ECUs to control vehicular functionalities by acquiring electronic signals from other components, as well as processing, and sending control signals. Some important ECUs are Brake Control Module, Engine Control Module, Tire-pressure Monitor Systems, and Inertial Measurement Units. Their functionalities are as follows. The Brake Control Module collects data from wheel-speed sensors and the brake system, as well as processes the data to determine whether or not to release braking pressure in real-time \cite{kassakian1996automotive}. The Engine Control Module controls fuel, air, and spark, as well as collects data from many sensors around the vehicle to ensure that all components are within a normal operating range \cite{kassakian1996automotive}. The Tire-pressure Monitor Systems collect data from sensors within tires and determine if the tire pressures are at ideal levels. The United States has legally required all vehicles to be equipped with Tire-pressure Monitor Systems since 2007 \cite{singh2009tire}, and the European Union issued the same regulation in 2012 \cite{regulation2009661}. The Inertial Measurement Units collect data from accelerometers, magnetometers, and gyroscopes and calculate the vehicle's velocity, acceleration, angular rate, and orientation. These calculations are pivotal for CAVs because they serve as inputs for running a safe automated driving system \cite{jitpakdee2008neural}. For example, a change in road gradient would change a CAV's angular rate and orientation, and the automated driving system may issue an adjustment in a vehicle's speed to maintain safe operations. CAVs involve a larger number of ECUs than a non-automated vehicle (SAE level 2 and below) because they possess many more sensors and require many more calculations to make autonomous decisions in driving. Readers may think of ECUs in CAVs as mini-computers, each carries out a specific role and collaborates with others to perform autonomous driving. It is common to see complex collaborations between ECUs \cite{koscher2010experimental}. Attack models and defense strategies on ECUs are discussed in section III-B. Communications between ECUs happen on Controller Area Networks, which will be discussed as follows.

\textbf{Controller Area Network (CAN)}. ECUs are typically connected through a CAN. In a vehicle, the CAN is a central network to connect ECUs together so that they can communicate with each other. A CAN bus is typically structured as a two-wire and half-duplex network system that can support high-speed communication \cite{hpl2002introduction}. The greatest benefits of CANs are the low amount of wiring and the ingenious prevention of message loss and message collision \cite{hpl2002introduction}. In CAVs, network packets are transmitted to all the nodes in the CAV network, and the packets do not contain an authentication field or source identification field \cite{thing2016autonomous}. Therefore, a compromised node can collect all data being transferred through the network and broadcast malicious data to other nodes, making the entire CAN vulnerable to cyberattacks. Attack models and defense strategies on CANs are discussed in section III-C.

\textbf{Sensors}. The following sensors are crucial to CAVs and are often found in most CAVs. All of the sensors discussed below will have their vulnerabilities discussed in Section III.
\begin{itemize}
    \item \textbf{Light Detection And Ranging (LiDAR)} are sensors that use light to measure the distance to surrounding objects. LiDAR sensors operate by sending light waves to probe the surrounding environment and make measurements based on reflected signals \cite{wandinger2005introduction}. The light beam's wavelength varies to suit the purpose and ranges from 10 micrometers (infrared light) to approximately 250 nanometers (ultraviolet light) \cite{wandinger2005introduction}. In CAVs, LiDAR is often used for obstacle detection to navigate safely through environments and is often implemented by rotating laser beams \cite{hecht2018lidar}. Data from LiDAR can be used by software embedded in ECUs to determine whether there are obstacles in the environment, as well as by autonomous emergency braking systems \cite{hulshof2013autonomous}. Attacks and defense techniques on LiDAR are described in section III-D.
    
    \item \textbf{Radio Detection and Ranging (Radar)} are sensors that send out electromagnetic waves in the radio or microwave domain to detect objects and measure their distance and speed by sensing the reflected signals. In CAVs, radars are useful in many applications. For example, short‑range radars enable blind-spot monitoring \cite{uselmann2004sonic}, lane-keeping assistance \cite{ishida2004development}, and parking aids \cite{reed2003vehicle}. Long‑range radars assist in automatic distance control \cite{chi1992automatic} and brake assistance \cite{breuer2007real}. Attacks and defense techniques on radars are described in section III-E.

    \item \textbf{Global Positioning System (GPS)} is a satellite-based navigation system that is funded and owned by the United States government, is operated and maintained by the United States Air Force \cite{team2014global}. It is a global navigation system that operates based on the satellites in the Earth's orbit that transmit high-frequency radio signals. The radio signals may be sensed by many devices such as smartphones and GPS receivers in CAVs. When GPS receivers find signals from three or more satellites, they can compute their locations. Since finding a route between two locations is necessary for autonomous driving, GPS signals are critical to CAVs. GPS receivers can operate without any communication channel such as wireless networks, but data from wireless networks can often enhance GPS receivers' accuracy \cite{twitchell2001gps}. Since GPS signals do not contain any data that can directly authenticate the source of signals, GPS receivers are vulnerable to jamming and spoofing attacks. These attacks and mitigation techniques are described in section III-F.
    
    \item \textbf{Cameras (image sensors)} are widely applied in CAVs. Autonomous and semi-autonomous vehicles (SAE level 2 and above) rely on cameras placed on many positions to acquire a 360-degree view around the vehicle. Cameras provide information for important autonomous tasks such as traffic sign recognition \cite{fairfield2011traffic,omachi2009traffic,levinson2011traffic} and lane detection \cite{sun2013robust,hillel2014recent}. Cameras can also be used to replace LiDAR for the task of object detection and for measuring distance at a lower cost, but they have poor performance under specific situations such as rain, fog, or snow \cite{wang2019pseudo}. Together with LiDAR and radars, cameras provide abundant and diverse data for autonomous driving. Attack models on cameras and  mitigation techniques are described in section III-G.
\end{itemize}

\textbf{Connection Mechanisms} in CAVs can be divided into vehicle-to-vehicle (V2V) and vehicle-to-infrastructure (V2I) networking. V2V communications help exchange data between nearby vehicles and can quickly provide additional information to the data already collected by a CAV regarding its surrounding environment. This additional data can lead to safer and more efficient autonomous driving. V2I communications help exchange data between CAVs and road infrastructures, which provide data about the bigger picture of the transportation system, such as smart traffic signs (without the need to do image recognition) and safety warnings in a large region \cite{chang2015estimated}. V2V communications often follow the Vehicular Ad-hoc NETworks (VANET) paradigm, where each vehicle acts as a network node and can independently interact with other nodes through a wireless connection \cite{watfa2010advances}. The wireless connections used are often dedicated short-range communications (DSRC) and cellular networks \cite{abboud2016interworking}. A well-known example of DSRC is the IEEE 802.11p Wireless Access in Vehicular Environments (WAVE) \cite{ieee2010ieee,baccelli2010ipv6}. WAVE is described in detail by Kenney (2011) \cite{kenney2011dedicated}. V2I communications are often achieved by using cellular networks, where Long-Term Evolution (LTE) is the current standard \cite{abboud2016interworking}. Attack models on V2V and V2I networks and defense techniques are discussed in section III-H.

\subsubsection{Classifications of Attack Models}
We can categorize the attack models, described in details in section III, by their access requirements and by their motives.

\textbf{Access Requirement:} attack models can be performed remotely (remote-access attacks) or can only be performed with physical access to CAV components (physical-access attacks).
\begin{itemize}
    \item \textbf{Remote-access attacks:} Attackers do not need to physically modify parts on CAVs or attach instruments to CAVs. Attacks can be launched from a distance, such as from another vehicle. Three common patterns for remote-access attacks are sending counterfeit data, blocking signals, and collecting confidential data. Examples of sending counterfeit data can be found in section III-D-Attack Model 1, section III-E-Attack Model 1, section III-F-Attack Model 1, and section III-G-Attack Model 2. Examples of attacks that block signals are described in section III-D-Attack Model 2, section III-E-Attack Model 2, section III-F-Attack Model 2, and section III-G-Attack Model 1. Examples of attacks that collect confidential data can be found in section III-H-Attack Model 1. Readers may refer to these sections for more details.
    
    \item \textbf{Physical-access attacks:} Attackers need to physically modify components on CAVs or attach instruments to CAVs. Examples of these attacks are reprogramming ECU (section III-A and section III-C-Attack Model 1) and falsifying input data (section III-B-Attack Model 1). Physical-access attacks are more difficult to carry out because attackers may be detected when tampering with CAVs.
\end{itemize}
For easier navigation through this paper, a summary of access requirements to perform attacks on the aforementioned CAV components is presented in Figure 3, annotated with the corresponding parts in section III for more details. CAN and ECUs can be targets for both remote-access and physical-access attacks.

\begin{figure*}
  \centering
  \captionsetup{justification=centering}
  \includegraphics[width=0.8\textwidth]{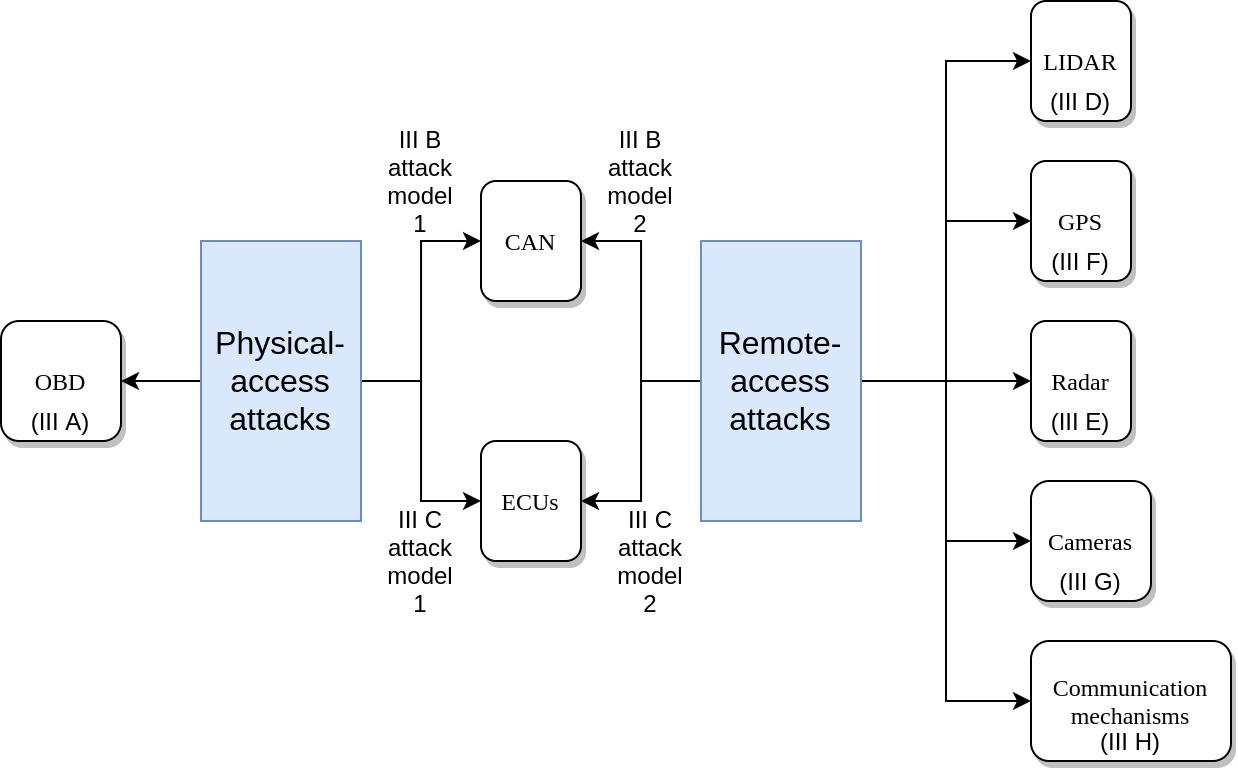}
  \caption{Access requirements to attack some CAV components}
\end{figure*}

\textbf{Attack Motivations:} Three common attack motivations are to interrupt (but without control) CAVs' operation, to control CAVs as attackers' wishes, or to steal information.
\begin{itemize}
    \item \textbf{Interrupting operations:} attackers aim to corrupt CAV components that are important for autonomous driving, thus making the autonomous driving mode unavailable on CAVs. These attacks are analogous to Denial-of-Service attacks on networks. Examples can be found in III-D-Attack Model 2, section III-E-Attack Model 2, section III-F-Attack Model 2, section III-G-Attack Model 1, and section III-H-Attack Model 2.
    \item \textbf{Gaining control over CAVs:} attackers gain sufficient control over CAVs so that they can alter the vehicles' movements, such as changing the vehicle's route, forcing emergency brake, and changing vehicle speed. Examples of these attacks can be found in section III-D-Attack Model 1, section III-E-Attack Model 1, section III-F-Attack Model 1, and section III-G-Attack Model 2.
    \item \textbf{Stealing information:} attackers' goal is to collect important and/or confidential information from CAVs. Collected information may be used for further attacks. Examples of this type of attack can be found in section III-B, section III-C, and section III-H-Attack Model 1.
\end{itemize}

For easier navigation through this paper, a summary of attack motives is presented in Figure 4, annotated with the corresponding parts in section III for more details.

\begin{figure*}
	\centering
	\captionsetup{justification=centering}
	\includegraphics[width=0.8\textwidth]{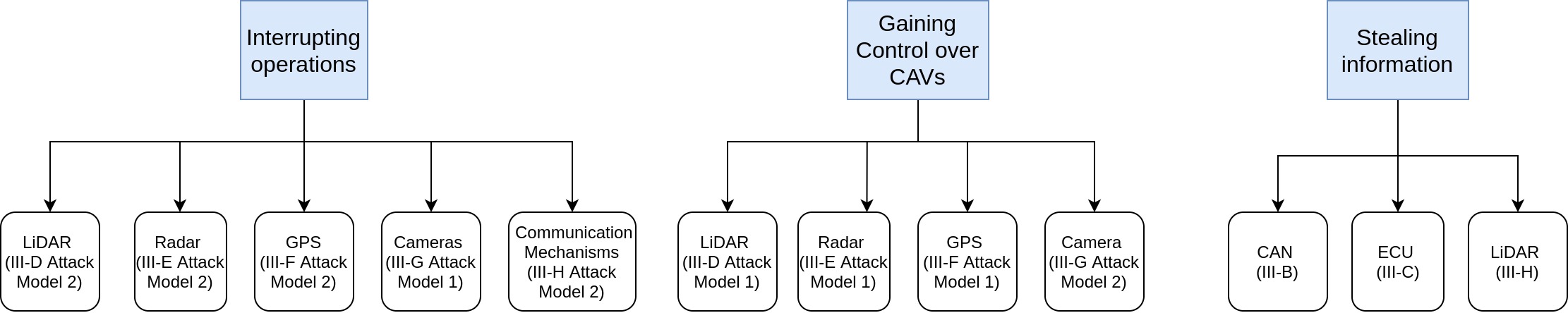}
	\caption{Attack motives}
\end{figure*}

Finally, for each attack model described in section III, we show its targeted CAV component, access requirement, and attack motives in Table II.
\begin{table*}[h]
	\caption{Classification of CAV attack models}
	\begin{center}
		
		\begin{tabular}{|>{\raggedright\arraybackslash}P{3cm} | P{3cm}| P{2cm} | P{5cm} |}
			\hline
			\multicolumn{1}{|>{\centering\arraybackslash}m{3cm}|}{\cellcolor{Gray}\textbf{Sub-section in section III}}
			& \multicolumn{1}{>{\centering\arraybackslash}m{3cm}|}{\cellcolor{Gray} \textbf{Targeted CAV component}}
			& \multicolumn{1}{>{\centering\arraybackslash}m{2cm}|}{\cellcolor{Gray} \textbf{Access requirement}}
			& \multicolumn{1}{>{\centering\arraybackslash}m{5cm}|}{\cellcolor{Gray} \textbf{Attack motives}}\\
			\hline

			\centering III-A & OBD & Physical & Interrupting operations, Gaining Control over CAVs, Stealing Information \\ 
			\hline
			\centering III-B Attack Model 1 & CAN & Physical & Interrupting operations, Gaining Control over CAVs, Stealing Information \\ 
			\hline
			\centering III-B Attack Model 2 & CAN & Remote & Interrupting operations, Gaining Control over CAVs, Stealing Information \\ 
			\hline
			\centering III-C Attack Model 1 & ECU & Physical & Interrupting operations, Gaining Control over CAVs, Stealing Information \\ 
			\hline
			\centering III-C Attack Model 2 & ECU & Remote & Interrupting operations, Gaining Control over CAVs, Stealing Information \\ 
			\hline
			\centering III-D Attack Model 1 & LiDAR & Remote & Gaining Control over CAVs\\ 
			\hline
			\centering III-D Attack Model 2 & LiDAR & Remote & Interrupting operations\\ 
			\hline
			\centering III-E Attack Model 1 & Radar & Remote & Gaining Control over CAVs\\ 
			\hline
			\centering III-E Attack Model 2 & Radar & Remote & Interrupting operations\\ 
			\hline
			\centering III-F Attack Model 1 & GPS & Remote & Gaining Control over CAVs\\ 
			\hline
			\centering III-F Attack Model 2 & GPS & Remote & Interrupting operations\\ 
			\hline
			\centering III-G Attack Model 1 & Camera & Remote & Gaining Control over CAVs\\ 
			\hline
			\centering III-G Attack Model 2 & Camera & Remote & Interrupting operations\\ 
			\hline
			\centering III-H Attack Model 1 & Connection Mechanism & Remote & Interrupting operations\\ 
			\hline
			\centering III-H Attack Model 2 & Connection Mechanism & Remote & Gaining Control over CAVs\\ 
			\hline
		\end{tabular}
	\end{center}
\end{table*}

\subsection{Taxonomy of Defenses}
In this section, We attempt to organize the defense techniques into categories that have certain patterns. Some defense categories are generally effective against certain types of attacks, but we suggest that readers study defense techniques for attacks on a case-by-case basis. For example, attacks that prevent sensors from receiving legitimate signals can usually be mitigated by the abundance of information, through planting multiple sensors or through acquiring additional information from V2V or v2I connections, but this does not hold for the case of GPS jamming (section III-F-Attack Model 2).

The categories for defense techniques are as follows.
\begin{itemize}
    \item \textbf{Anomaly-based Intrusion Detection System (IDS)} is a method that is designed to detect unauthorized access or counterfeit data. IDS methods generally look to detect abnormal data from the signals or side-channel information. For example, Cho et al. (2016) \cite{cho2016fingerprinting} proposed Clock-based IDS (CIDS), which measures the clock skew of ECUs (the phenomenon in which the clock signal arrives at different ECU at slightly different times) and uses this information to fingerprint the ECUs. The fingerprints are then used to detect intrusions by checking for any abnormal shifts in the clock skews. IDS methods are generally applicable in attack models that rely on sending counterfeit signals, such as CAN attacks (section III-B), LiDAR spoofing (section III-D-Attack Model 1), radar spoofing (section III-E-Attack Model 1), and GPS spoofing (section III-F-Attack Model 1). However, they are not effective to defend against adversarial image attacks on cameras (section III-G-Attack Model 2).
    
    \item \textbf{The abundance of information} can be achieved by getting information from other CAVs and infrastructures or by setting up abundant information within a CAV. This is good not only for defending against cyberattacks but also for increasing confidence in autonomous driving. This category of defense strategy is generally effective against Denial-of-Service type of attacks, such as LiDAR jamming (section III-D-Attack Model 2), radar jamming (section III-D-Attack Model 2), and camera blinding (section III-G-Attack Model 1). For example, by using multiple LiDAR sensors with different wavelengths, a CAV is protected from attackers who send high-power light beams to blind LiDAR sensors. However, this method is not effective against GPS jamming (section III-F-Attack Model 2). A major drawback of this defense category is that placing abundant components on CAVs is expensive.
    
    \item \textbf{Encryption methods} can be applied to defend against attacks that abuse components that lack authentication methods, such as CANs and signals for sensors. Many encryption methods have been published to secure CAN and sensor signals, such as those in \cite{lin2012cyber,nilsson2008framework,van2011canauth,halabi2018lightweight} (they will be discussed in detail in section III). However, an encryption method is not applicable for GPS receivers because it is too expensive to modify the satellites so that they can send encrypted radio signals. This makes defending against GPS jamming attacks a difficult task (section III-F-Attack Model 2).
\end{itemize}

Some defense techniques are unique and do not fall into any of these categories, which is why we suggest that readers study defense techniques on a case-by-case basis.

\section{Existing attacks and their countermeasures}
One compromised component of CAVs may allow attackers to compromise other components, other CAVs, and infrastructures, thus forming a sequence of attacks. From our reading of the literature, we have come up with possible attack sequences that attackers may perform and present them in Figure 5. The attack sequences plotted in Figure 5 are:

\begin{itemize}
    \item Sequence with label (1): Attackers gain physical access to OBD ports, which gives them access to CANs and subsequently access to ECUs.
    \item Sequence with label (2): Attackers compromise LiDAR, radar, GPS, or cameras and send adversarial information to ECUs.
    \item Sequence with label (3): Attackers compromise telematics ECUs (ones that have access to communication channels such as VANET, Bluetooth, and DSRC). Attackers can then send adversarial information through the CAN to other ECUs.
    \item Sequence with label (4): Attackers can send adversarial information from their CAVs or CAVs that have been compromised.
\end{itemize}

The possibility that attackers may compromise one CAV component after another means that in order to enhance the security of CAVs, manufacturers should enhance the security of all CAV components.

\begin{figure*}
  \centering
  \captionsetup{justification=centering}
  \includegraphics[width=\textwidth]{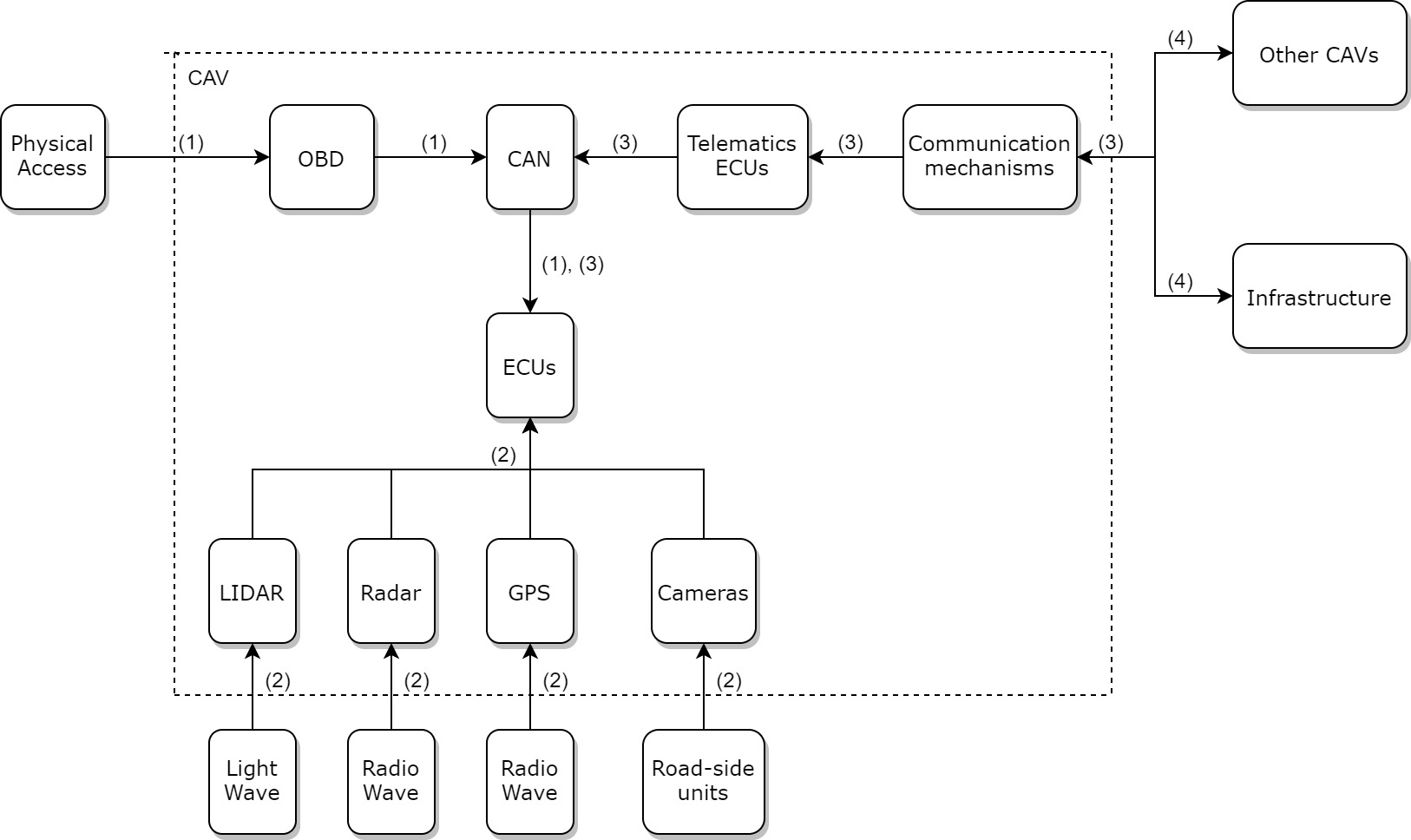}
  \caption{Possible attack sequences on the components of CAVs}
\end{figure*}

Each of the following subsections describes attack model(s) for a specific CAV component. In each subsection, we describe the attack models, the security requirements for defense techniques against the attack models, existing defense techniques in the literature and whether they meet the security requirements, and challenges for defending against the attack models. We number the attack models for easier reference to section II and include short descriptions as the names of the attack models. For example, in section III-B, \textbf{Attack Model~1 - CAN access through OBD} would be referred as "III-B Attack Model 1" in section II and this attack models happen through physical access to the OBD port.

\subsection{Attacks on OBD}
\textbf{Attack Model 1 - malicious OBD devices:} The OBD port is an open gateway for many attacks to other CAV components because the port usually does not encrypt data or control access. Since the OBD port itself has no capability of remote connection, attackers would need physical access to the OBD port to perform these attacks. Some devices that are plugged into the OBD port can transfer data to a computer through wired or wireless connections. Some of these devices were made by car manufacturers for diagnostic purposes and firmware updates. Some examples of these devices are Honda’s HDS, Toyota's TIS, Nissan’s Consult 3, and Ford’s VCM \cite{checkoway2011comprehensive}. Some other devices are developed by third-party companies to connect vehicles to smartphones (i.e. self-diagnostic purpose), such as Telia Sense \cite{uhlir2017practial} and AutoPi \cite{zamfir2019automotive}. Studies have been done to assess the feasibility of using these third-party devices to perform a meaningful attack. Marstorp and Lindstr{\"o}m (2017) \cite{marstorp2017security} found that Telia Sense is a well secured system whereas Christensen and Dannberg (2019) \cite{christensen2019ethical} successfully performed a man-in-the-middle attack, where they intercepted data to and from the AutoPi Cloud interface. After gaining access to the OBD port, attackers can interrogate information about the CAV, controlling key components (such as warning light \cite{ishtiaq2010security}, windows lift, airbag control system, horn \cite{koscher2010experimental}), and injecting codes to ECUs \cite{miller2013adventures}. Koscher et al. (2011) \cite{koscher2010experimental} successfully performed one such attack on a running vehicle by using a self-written program named CARSHARK to compromise many components on the vehicle. In \cite{woo2014practical}, it was shown that if attackers can trick drivers into downloading a malicious self-diagnostic application on their smartphones, attackers can transmit an ECU-controlling data frame through the OBD device to the vehicle's ECUs.

\textbf{Criteria for Defense Strategies:} Defense may take place at the OBD level, or at CAN and ECUs level. Criteria for defense strategies at CAN and ECUs are discussed in their respective subsections (III-B and III-C). Here, we discuss the criteria for securing the OBD port. Based on our understanding of the attack model and study on the defense techniques, we suggest the following criteria.
\begin{itemize}
	\item Authenticity of OBD devices: before being granted access to a CAV's data, OBD devices must come from a trusted manufacturer 
	\item Integrity of OBD devices: they must be provable that they have not been compromised or corrupted after their creation.
	\item Privacy of OBD devices: any information gained from the OBD port is intelligible only to the device's intended party.
	\item The authentication process of OBD devices should be efficient to not cause significant delay to users.
\end{itemize}

\textbf{Existing Defense Strategies:} Unfortunately, we have not found any significant method in the literature to secure the OBD port and to detect malicious devices. However, defense strategies for CAN and ECUs are abundant. Defense strategies to detect abnormal activities from OBD ports can be implemented in CANs and ECUs and are described in their respective subsections. Fowler et al. (2017) \cite{fowler2017towards} proposed using hardware-in-the-loop (HIL) equipment to collect and simulate data on attacks through an OBD port. The HIL technique is not a defense layer on OBD ports but provide a virtual environment for further testing attack and defense mechanisms for the OBD port.

\textbf{Challenges for defending against this attack model:} Because OBD ports are commonly used by car manufacturers for diagnostics and firmware updates, and by other companies for data collection purposes, it is difficult to distinguish legitimate from malicious OBD devices. No study has proposed putting a layer of defense in the OBD port and this remains a challenging problem.

\subsection{Attacks on CAN}
\textbf{Attack Model 1 - CAN access through OBD:} Since CAN protocols generally do not support encrypting messages for authentication and confidentiality \cite{matsumoto2012method}, attackers can perform three types of attack as follows. 
\begin{itemize}
    \item Eavesdrop: CAN messages can be observed from the OBD port \cite{matsumoto2012method}.
    \item Replay attack and unauthorized data transmission \cite{lin2012cyber}: Once attackers have observed all messages transmitted on the CAN bus, they can easily impersonate an ECU and transmit counterfeit messages through OBD ports.
    \item Denial of Service attack \cite{matsumoto2012method}. Attackers can send many messages with high priority through OBD ports and prevent CAN from processing other messages on the CAN bus.
\end{itemize}
It is important to note that all of these attacks require physical access to the OBD port.

\textbf{Criteria for Defense Strategies against Attack Model 1:} Based on our understanding of the attack model and study on the defense techniques, we suggest the following criteria.
\begin{itemize}
	\item Confidentiality/privacy of CAN messages: CAN messages should be readable by only the intended receiving ECUs. A message sent through CAN is received by all ECUs connected to that CAN and each ECU decides whether to use it by checking the identifier of the message \cite{hoppe2008security}. This may allow attackers to conclude private information such as driving behavior or the state of the vehicle.
	\item Authenticity of CAN messages: CAN messages should only be sent from verified ECUs connect to that CAN to prevent unauthorized data transmission and Denial of Service attacks.
	\item Low requirement for computing resources: the authentication and verification process of CAN messages should be done efficiently to ensure real-time performance of the entire CAV system.
\end{itemize}

\textbf{Existing Defense Strategies against Attack Model 1:} Wolf et al. (2006) \cite{wolf2006secure} proposed a secure CAN protocol that achieved authenticity and confidentiality by using Symmetric Key Encryption and Public Key Encryption. Lin et al. (2012) \cite{lin2012cyber} and Nilsson et al. (2008) \cite{nilsson2008efficient} also proposed to achieve authenticity by using the Message Authentication Code (MAC) method. Herrewege et al. (2011) \cite{van2011canauth} proposed an authentication protocol named CANAuth, which also uses MAC for authentication but utilizes an out-of-band channel to send more authentication data in a real-time environment. Even with the out-of-band channel, Herrewege et al. acknowledge that public-key cryptography is not viable due to its large key size requirement. Matsumoto et al (2012). \cite{matsumoto2012method} criticized that the aforementioned cryptographic methods suffer from key management issues (e.g. leakage of secret keys) and may not be fast enough to achieve real-time response in a moving vehicle. Halabi and Artail (2018) \cite{halabi2018lightweight} aimed to solve these problems by proposing lightweight symmetrical encryption where the keys are generated based on a CAN frame's payload and the previous key.

IDS-based defense techniques are popular. Müter et al. (2011) \cite{muter2011entropy} proposed calculating the entropy of a CAN bus during normal activities; significant deviations in entropy are then used to detect attacks. Similarly, Miller et al. (2014) \cite{miller2014survey} proposed using a small device that connects to the OBD port, collects traffic data, and detects abnormal traffic patterns with machine learning. When the device detects an attack, it stops the circuit on the CAN bus and disables all CAN messages. Matsumoto et al. (2012) \cite{matsumoto2012method} proposed a solution where all ECUs attempt to detect unauthorized messages by monitoring all messages being transmitted on the CAN bus. For each ECU, a flag is implemented within the CAN controller that would indicate whether the ECU is trying to send a message. Then, the timing of the flag being switched is measured to detect unauthorized messages. Shin and Cho (2017) \cite{shin2017fingerprinting, cho2016fingerprinting} proposed a similar method, where they measured the intervals of periodic CAN messages and used these measurements to detect abnormal messages. Gmiden et al. (2016) \cite{gmiden2016intrusion} criticized that Matsumoto's method requires modifications to each ECU and thus is an expensive solution. Gmiden et al. then proposed an IDS that checks the identity of each ECU that send CAN messages and calculates the time since the last message from the same ECU was observed. If the new time interval is significantly shorter than the previous time intervals from the same CAN ID, an alert of attack is raised. Tyree et al. (2018) \cite{tyree2018exploiting} proposed an IDS that uses the correlations between ECU messages to estimate the state of the vehicle. A sudden change to an ECU's messages would raise an alert that the ECU is compromised. If an attacker successfully compromises many ECUs, the sudden change in the state of the vehicle is used to detect the attack. Many more types of IDS for CAN can be found in Tomlinson's survey paper about this topic \cite{tomlinson2018towards}. Siddiqui et al. (2017) \cite{siddiqui2017secure} proposed a hardware-based framework that implements mutual authentication and encryption over the CAN Bus.

\textbf{Challenges for defending against Attack Model 1:} Requirements for a good defense method against Attack Model 1 are real-time response, accuracy, and low degree of modification needed on the vehicle. Even though many defense mechanisms have been proposed in the literature \cite{tomlinson2018towards}, there also exist criticisms to most of them as discussed in the previous two paragraphs. It is difficult to determine the best defense strategies to implement in an operating CAV. Therefore, there is a need for comparative studies that are performed on moving vehicles to serve as recommendations for CAV manufacturers. In addition, CAN standards may carry legacy and thus may not have the capacity to accommodate the computing demand and communication constraints to these innovative solutions.

\textbf{Attack Model 2 - CAN access from telematics ECUs:} One may attack a CAN by compromising a member ECU first. Some ECUs could be compromised without access to the CAN because they have other access points through connection mechanisms such as Bluetooth and cellular networking. The compromised ECU can then send authenticated messages that can bypass the cryptographic and IDS-based defense mechanisms discussed in Attack Model 1. More details about how telematics ECUs are compromised and studies that have implemented this attack model are presented in subsection C.

\textbf{Criteria for Defense Strategies against Attack Model 2:} Defense strategies may be implemented on the CAN to detect abnormal ECU behaviors, or may be implemented on telematics ECUs to prevent compromise. Here, we discuss the second approach and leave the first approach for section III-C. The criteria for a defense strategy to be implemented on CAN are:
\begin{itemize}
	\item Enforce integrity of messages sent through CAN under the possibility that a connected ECU is compromised (and thus any authenticity test is invalid). 
	\item Efficient verification process to not interrupt system-wide service.
\end{itemize}

\textbf{Existing Defense Strategies against Attack Model 2} When attackers have successfully compromised an ECU, they may be able to send encrypted and authenticated messages through the CAN. Therefore, cryptographic methods discussed in Attack Model 1 such as Symmetric Key Encryption, Public Key Encryption, and MAC may not be able to detect the attack. Some IDS-based defense techniques would not be useful either. For example, Matsumoto's method \cite{matsumoto2012method} and Gmiden's method \cite{gmiden2016intrusion} may not work because messages being sent from the compromised ECU would not create any timing abnormality. Other IDS-based approaches, such as Müter's \cite{muter2011entropy}, Miller's \cite{miller2014survey}, and Tyree's \cite{tyree2018exploiting}, may work because they attempt to model the traffic patterns and state of the vehicles, thus we speculate that they may acquire essential information to detect abnormal messages from the compromised ECU. A probably better strategy of defense is to secure the telematics ECUs. Securing telematics ECUs are discussed in the next subsection about ECUs.

\textbf{Challenges for defending against Attack Model 2:} As previously discussed, there are two approaches to defend against Attack Model 2: securing telematics ECUs \cite{nilsson2008framework,seshadri2006scuba} and using IDS-based algorithms to detect abnormal traffic patterns from the compromised telematics ECUs \cite{muter2011entropy,miller2014survey,tyree2018exploiting}. Implementing both approaches in a CAV would increase the security against this attack model. However, it is still unclear how effective these approaches will be and how difficult they are to be implemented. 

\subsection{Attacks on Electronic Control Units}
\textbf{Attack Model 1 - ECU access through CAN:} Attackers compromise ECUs through their access to CAN. As previously discussed, after gaining access to CAN through the OBD port or telematics ECUs, attackers may compromise other ECUs on the CAN. Examples of attack techniques are falsifying input data \cite{koscher2010experimental}, code injection and reprogramming ECUs \cite{prathap2013penetration,hoppe2008security}.

\textbf{Existing Defense Strategies against Attack Model 1:} have been discussed in sections regarding OBD and CAN (III-A and III-B).

\textbf{Attack Model 2 - ECU access through connection mechanisms:} Attackers compromise a telematics ECUs through their connection mechanisms. We have found two examples of this attack model in the literature. First, Checkoway et al. (2011) \cite{checkoway2011comprehensive} were able to get remote code executions on a telematics unit of a vehicle through Bluetooth and long-range wireless connection (chapters 4.3 and 4.4 on the cited paper, respectively). The authors achieved this result by extracting the ECUs' firmware and using disassembly (a computer program that decompiles machine code into assembly language) to reverse-engineer the code. After assessing the firmware of the ECU that is responsible for Bluetooth connections, the authors hypothesized that if attackers can pair their smartphones with the Bluetooth ECU, they can compromise the ECU by sending malicious code through their smartphones. For example, after re-engineering the operating system of the ECU that is responsible for handling Bluetooth connections, Checkoway et al. found over 20 insecure calls to \textit{strcpy}, one of which would allow them to copy data to the stack and thus to execute any code on the ECU. Second, Nilsson et al. (2008) \cite{nilsson2008framework} described another pathway for attacking telematics ECUs as follows. Many CAV manufacturers are performing firmware updates over the air (FOTA) for ECUs \cite{nilsson2008creating}. The FOTA process works as follows. The firmware is first downloaded over a wireless network connection to a trusted station, then transferred to the vehicle, and finally transferred to the ECUs. The firmware transferring process can be secured by using protocols described in \cite{mahmud2005secure, nilsson2008secure}, but the installation process is not secured. Therefore, the downloaded firmware is vulnerable to susceptible to adversarial modification by a time-of-check-to-time-of-use attack (TOCTTOU), as described in \cite{mulliner2012read}. The TOCTTOU attack works as follows. Given the benign firmware update \textit{File B} that is expected by the check-install code. The attacker also prepares malicious \textit{File M} and constructs a storage device that can observe the read requests to \textit{File B}. For the first access to \textit{File B}, the mass storage device serves the legitimate \textit{File B}. This first access is likely to serve the purpose of calculating and comparing the cryptographic hashes. After the verification process succeeds, the storage device serves the malicious \textit{File M} for the installation phase. The attack succeeds if the check code verifies the benign file \textit{File B} and then install the malicious \textit{File M} for the ECU's firmware update.

\textbf{Criteria for Defense Strategies against Attack Model 2:} Based on our understanding of the attack model and study on the defense techniques, we suggest the following criteria.
\begin{itemize}
	\item Robust code on ECUs' firmware to avoid code injection.
	\item Restrict access to connect to telematics ECUs, i.e., only accept connections from trusted and authenticated sources.
	\item Robust firmware update protocols for ECUs that assure integrity and authenticity of the firmware updates.
\end{itemize}

\textbf{Existing Defense Strategies against Attack Model 2:} Checkoway et al. \cite{checkoway2011comprehensive} did not discuss specific defense strategy against their attack model through Bluetooth and wireless connection, but did mention that robust code and firmware update protocols for ECUs are necessary. Seshadri et al. (2006) \cite{seshadri2006scuba} and Nilsson et al. (2008) \cite{nilsson2008framework} proposed a protocol to secure the ECUs' FOTA. In \cite{nilsson2008framework}, the secure protocol can be summarized as follows. First, the trusted station generates a random value and combines it with fragments of the firmware update to create a chain. The chain is then hashed repeatedly and the final hash value serves as the verification code. Next, the firmware, the random value, and the verification code are transferred to the vehicle over a secure channel and the hashing process is performed again to validate the integrity of the firmware. In \cite{seshadri2006scuba}, the authors proposed Indisputable Code Execution (ICE), which is a protocol to securely execute codes on a network node from a trusted station. ICE consists of three steps: checking the integrity of the firmware update code, setting up an environment in which once the firmware update is executed, no other code is allowed to be executed, and executing the firmware update within the safe environment.

\textbf{Challenges for defending against Attack Model 2:} We have not found any study that implements the frameworks described in \cite{nilsson2008framework} and \cite{seshadri2006scuba} to secure FOTA. A study to validate these frameworks in a realistic CAV environment is still needed as these publications only presented theoretical frameworks.

\subsection{Attacks on LiDAR}
\textbf{Attack Model 1 - LiDAR spoofing:} An attacker can record legitimate signals sent from a LiDAR sensor and relay the signals to another LiDAR sensor of the same CAV to make real objects appear closer or further than their actual locations. Another variation of this attack model is when an attacker creates counterfeit signals that represent an object and inject the counterfeit signals into a LiDAR sensor. Both of these variations have been successfully demonstrated by Petit et al. (2015) \cite{petit2015remote} at a low financial cost. The authors performed the spoofing attack as follows. The attacker uses two transceivers \textit{B} and \textit{C}. The output of \textit{B} is a voltage signal that corresponds to the intensity of the pulse sent by the LiDAR device being attacked. The output of \textit{B} is sent to \textit{C}, which in turn emits a pulse to the LiDAR device. The total cost of these two transceivers was 49.9 US Dollars! Despite the low cost, the authors managed to make the vehicle’s ECU (one that receives input from a LiDAR sensor) think that it is approaching a large object and initiate emergency brake. The second attack variation was more difficult to perform, as it requires the attacker to send the counterfeit signals within a small window after a LiDAR sensor sends its signal. Shin et al. (2017) \cite{shin2017illusion} performed this attack variation on a stand-alone LiDAR sensor, Velodyne’s VLP-16. Cao et al. (2019) \cite{cao2019adversarial} concluded from their experiments that the machine learning-based object detection process made it difficult to perform a LiDAR spoofing attack. Nevertheless, the authors formulated an optimization model for the process of generating counterfeit inputs. They performed a case study on the Baidu Apollo's software module and was able to force an emergency brake, reducing the vehicle's speeding from 43 km/h to 0 in a second. The authors claimed that their attack model can have a success rate of 75\%.

\textbf{Criteria for Defense Strategies against Attack Model 1:} Based on our understanding of the attack model and study on the defense techniques, we suggest the following criteria.
\begin{itemize}
	\item Low cost: A trivial solution for defending against this attack model is to have redundant LiDAR devices on the vehicle to make it more difficult for attackers to spoof all devices simultaneously. However, the cost for this solution is proportional to the number of redundant LiDAR devices and thus this solution may not be appealing for manufacturers.
	\item High immediacy: The solution should not take too long to detect a spoofing attack. This also implies that the solution should be computationally efficient.
	\item Signal filter: a solution that can detect spoofing attacks may be sufficient, but a solution that can filter out legitimate signals among adversarial signals would be even more appealing.
\end{itemize}

\textbf{Existing Defense Strategies against Attack Model 1:} Shin et al. \cite{shin2017illusion} proposed a few defense strategies such as using multiple sensors having overlapping views, reducing the signal-receiving angle, transmitting pulses in random directions, and randomizing the pulses' waveforms. Nevertheless, the authors also pointed out that these defense strategies do not match all the aforementioned criteria. Using multiple sensors is expensive. Reducing the signal-receiving angle is also expensive because it requires more LiDAR devices to cover the entire space around the vehicle. Transmitting pulses in random directions is feasible and inexpensive, but does not have good immediacy because the LiDAR device would have to send many unused pulses. Randomizing the pulses' waveforms and rejecting pulses different from the transmitted one is probably the most appealing solution thanks to its low cost and high immediacy. Approaches of this type have also been studied intensively and applied for military radars \cite{pace2009detecting}. In a 2016 study on LiDAR spoofing on Unmanned Aerial Vehicle (UAV), Davidson et al. (2016) \cite{davidson2016controlling} proposed using LiDAR data in previous frames to formulate a momentum model that aims to detect adversarial inputs. The model utilizes the random sample consensus (RANSAC) method and works as follows. Let $v_{diff}=(dx_1,dy_1),(dx_2, dy_2), ...$ be the vector of motion that contain features from LiDAR object detection. RANSAC randomly samples k features and forms a hypothesis for each of them. The hypothesis $h_j$ for vector $(dx_j,dy_j)$ is the ground truth motion for vector $(dx_j,dy_j)$. Then, we let all other features vote for each of these k hypotheses. For a feature motion $(dx_j,dy_j)$ to vote for a hypothesis $h_j$, the two motion vectors need to be similar such that $|(dx_i-dx_j,dy_i-dy_j)|_1 < threshold$. The RANSAC method performs many realizations of this process and then picks the hypothesis with the highest vote to be the final hypothesis for the frame, and the corresponding features to be the ground truth of the frame. The shortcoming of this solution, which the authors also acknowledged in the paper, is that it would take some time to build up the weights for the model and would require high computational power. Thus, the solution is not immediate. In 2018, Matsumura et al. \cite{matsumura2018secure} proposed a mitigation technique that embeds the authentication data onto the light wave itself. The fingerprinting is obtained by modulating LiDAR's laser light with information from a cryptographic device, such as an AES encryption circuit. This defense strategy is interesting because it is cost-effective to implement and the authors claimed that attackers cannot make a distance-decreasing attack larger than 30 cm. It is important to note that Cao's study (2019) \cite{cao2019adversarial} on an attack model was published after Matsumura's defense strategy (2018) \cite{matsumura2018secure} but did not discuss this defense strategy any other countermeasure. In 2020, Porter et al. \cite{} proposed adding dynamic watermarking to LiDAR signals to validate measurements. This solution has the potential to satisfy all the three aforementioned criteria.

\textbf{Challenges for defending against Attack Model 1:} Cao's attack model may be considered state-of-the-art for its newness and effectiveness. Matsumura's countermeasure is also novel and recent \cite{matsumura2018secure}. It will be an interesting study to implement Matsumura's strategy against Cao's attack model \cite{cao2019adversarial}. Davidson's proposed method for UAV \cite{davidson2016controlling} may also be worth an experiment on CAVs. Porter et al.'s solution shows good potential and will be interesting for the community to discuss.

\textbf{Attack Model 2 - LiDAR jamming:} Attackers aim to perform a Denial-of-Service attack by sending out light with the same wavelength but with higher intensity and effectively preventing the sensor from acquiring the legitimate light wave. This technique has been used by civilians who aim to avoid speeding tickets by jamming police's speed gun (a LiDAR device) \cite{foxnews}. In the context of CAV attacks, Stottelaar (2015) \cite{stottelaar2015practical} successfully performed a jamming attack on a LiDAR sensor (Ibeo Lux3) and argued that such an attack on a CAV's LiDAR sensor is possible. We have not found any other study or experiment on LiDAR jamming in the literature. Nevertheless, the attack process, as described in detail in \cite{stottelaar2015practical}, is relatively straightforward and should not require much training to replicate.

\textbf{Criteria for Defense Strategies against Attack Model 2:} Based on our understanding of the attack model and study on the defense techniques, we suggest the following criteria.
\begin{itemize}
	\item Low cost: The solution should not require expensive modification to the vehicle
	\item High immediacy: The solution should not take too long to detect a jamming attack.
	\item Signal filter: a solution should be able to filter out legitimate signals among jamming signals.
\end{itemize}

\textbf{Existing Defense Strategies against Attack Model 2:} Stottelaar \cite{stottelaar2015practical} suggested several countermeasures such as using V2V communications to gather additional information, changing the wavelength frequently, using multiple LiDAR sensors with different wavelengths, and shortening the ping period (the time window that a sensor waits for the signal to come back). However, these countermeasures all have disadvantages. Vehicle-to-vehicle communication may not always be available. Using multiple LiDAR devices is expensive. The shortened ping period makes a device prone to errors. Changing wavelength frequently may not be effective against attackers who can follow a CAV for a while, as acknowledged by the author. Wang et al. (2015) proposed a novel LiDAR scheme called pseudo-random modulation (PMQSL) quantum secured LiDAR \cite{wang2015pseudorandom}. The PMQSL scheme is based on random modulation technique. The random modulation is a technique in the time domain, which is a typical way to recover the weak signal buried in random noise. The transmitting signal is modulated by the digital pulse codes, usually consisting of on and off. The $n^{th}$ order M-sequence $a_i$ with elements 1 or 0 is generated by a set of n-stage shift registers. The laser is pulse-position modulated by an electro-optic modulator. These pulses are further randomly modulated to create the horizontal, diagonal, vertical, and anti-diagonal polarization states of the photon through a polarization modulated model. When there is no jamming attack, four different distances corresponding to the four measured polarizations have very small error rates. In the presence of jamming attacks, the four distances have considerable error in the received polarization. The increase in error allows the system to determine that the LiDAR device was being jammed. The authors claimed that this LiDAR scheme can efficiently detect a jamming attack, but cannot filter out legitimate signals among jamming signals.

\textbf{Challenges for defending against Attack Model 2:} We have not found any study or experiment that demonstrates LiDAR jamming on a moving autonomous vehicle, as in Attack Model 1. Such a study will be interesting since we can observe the real effectiveness of Attack Model 2. Besides, defense strategies proposed in \cite{wang2015pseudorandom} and \cite{nouri2017target} need to be tested for CAVs because they were not specifically developed CAVs.

\subsection{Attacks on radar}
\textbf{Attack Model 1 - Radar spoofing:} Attackers replicate and rebroadcast radar signals to inject distorted data to the sensor. A common tool to perform this attack model is Digital radio frequency memory (DRFM), which is an electronic method to store radio frequency and microwave signals by using high-speed sampling and digital memory \cite{roome1990digital}. The phase of stored signals can then be modified and the signals are then re-broadcasted to the radar sensor. The falsified signals can then cause incorrect calculations of distance to surrounding objects. Chauhan (2014) \cite{chauhan2014platform} experimented with this attack model on a radar device (Ettus Research USRP N210) and managed to make an object from a 121-meter distance appear at a 15-meter distance. Yan et al. (2016) \cite{yan2016can} discussed the same idea of the attack but unfortunately, they did not have the resource to implement the attack. Instead, they attempted to inject counterfeit signals to a radar sensor on a Tesla Model S. The attempt was not successful because the sensor has a low ratio of working time over idle time, which makes it difficult to inject signals at the precise time slot.

\textbf{Criteria for Defense Strategies against Attack Model 1:} Based on our understanding of the attack model and study on the defense techniques, we suggest the following criteria.
\begin{itemize}
	\item Attack Detection: The solution should be able to detect a radar spoofing attack in a timely manner.
	\item Signal Filter: The solution should be able to filter out the attack signals and derives accurate distance measurements.
	\item Consistency: The solution should be able to achieve the previous two criteria under many circumstances and over a long period of time.
	\item Non-disruptivity: The solution should not affect other services of a vehicle.
\end{itemize}

\textbf{Existing Defense Strategies against Attack Model 1:} A novel approach, called physical challenge-response authentication (PyCRA) (2015) \cite{shoukry2015pycra}, inspects the surrounding environment by sending randomized probing signals, called challenging signals. PyCRA shuts down the actual sensing signals at random times and assumes that attackers cannot detect challenging signals immediately. Under that assumption, PyCRA can detect malicious signals by determining if they are higher than a noise threshold during a period with the Chi-square test. Kapoor et al. (2018) \cite{kapoor2018detecting} criticized PyCRA that PyCRA may severely affect the safety-critical CAV components, such as adaptive cruise control and collision warning, because they are shut down at random times. Another shortcoming of PyCRA is that after the first 30 seconds following an attack, The derived distance is continuously longer than the actual distance \cite{kapoor2018detecting}, thus PyCRA falls short of the Consistency criteria. Dutta et al. (2017) \cite{dutta2017estimation} attempted to address PyCRA's problems with consistency and non-disruptivity by introducing the Challenge Response Authentication method (CRA). CRA works by applying the recursive least square method to provide the estimated distance by minimizing the sum of square of errors, which is defined as the difference between the predicted distance and the actual distance. According to Dutta et al., CRA may satisfy all four aforementioned criteria. However, Kapoor et al. criticized that CRA may not be effective in practice because it relies on the assumption that the actual distance is known \cite{kapoor2018detecting}. Kapoor et al. proposed a new method called Spatio-Temporal Challenge-Response (STCR). STCR uses the same idea as PyCRA, but instead of shutting down sensing signals, it transmits challenging signals randomized directions. Reflected challenging signals can be used to identify directions that reflect malicious signals, then excludes the untrustworthy directions when measuring the surrounding environment. According to \cite{kapoor2018detecting}, STCR is able to detect attacks and measure the actual distance consistently and in a timely manner.

\textbf{Challenges for defending against Attack Model 1:} Yan et al. \cite{yan2016can} failed to apply the DRFM technique to inject counterfeit signals to a radar sensor because the sensor has a low ratio of working time over idle time. However, it was unclear from their publication whether this is the characteristic that all CAVs' radar sensors share, and whether an attacker can find a way to overcome this problem. Further experiments are needed to answer this question. Besides, Kapoor et al.'s defense method \cite{kapoor2018detecting} is the state-of-the-art technique but has not been validated in an experiment.

\textbf{Attack Model 2 - Radar jamming:} This attack model can also be carried out by using DRFM, but instead of modifying phase, attackers can modify frequency and amplitude of the stored signals before rebroadcasting to the radar sensors. The falsified signals can make the radar sensors fail to detect the object, at which the jamming device is located. We have not found any publication that experimented with this attack model on CAVs. However, this attack model is widely used by manned and unmanned aerial vehicles (UAV) to hide themselves from radar detection \cite{buehler2014airborne,stott1994digital,Lothes1990jammingbook}. Since these attacks have only been applied to UAVs and not CAVs, we are not certain about their feasibility on CAVs and thus cannot make claims about criteria for defense strategies.

\textbf{Existing Defense Strategies against Attack Model 2:} Similar to the attack model, defense strategies are widely studied for UAVs, but none has been discussed for CAVs. To defend against this attack model, one can attempt to separate the legitimate signals from the counterfeit signals. The specific methods are described in \cite{greco2005combined,greco2008radar,lu2010anti,nouri2017target}.  

\textbf{Challenges for defending against Attack Model 2:} Further studies on both attacks and defense techniques are needed to investigate whether the attacks are feasible on CAVs and whether the defense techniques are also effective on CAVs.

\subsection{Attacks on GPS}

\textbf{Attack Model 1 - GPS Spoofing:} An attacker broadcasts incorrect, but realistic GPS signals to mislead GPS receivers on CAVs. This is also known as a GPS Spoofing attack. In this attack model, attackers begin by broadcasting signals that are identical to the satellites' legitimate signals. The attackers then gradually increase the power of his signals and gradually deviate their GPS signals from the target's true location. GPS receivers are often configured to make use of signals with the strongest magnitudes \cite{krasner2000method}. Therefore, once the counterfeit signal is stronger than the legitimate satellite signal, GPS devices would choose to process the counterfeit signal. Tippenhauer et al. (2011) described in detail how to perform a GPS Spoofing attack and the requirements for a successful attack \cite{tippenhauer2011requirements}. They found that an attacker must be able to calculate the distance from himself to the victim with an error of at most 22.5 meters. Whether this condition could be met on a moving CAV is still unclear. We have not found any successful GPS spoofing attack on CAV published in the literature. However, attacks on other transportation means are available. For instance, in 2014, Psiaki et al. \cite{psiaki2014gnss} successfully spoofed GPS signals to a superyacht's and reported counterfeit locations to the crew. The crew then attempted to correct the course, only to deviate from the correct course. Shepard et al. (2012) \cite{shepard2012evaluation} used a civilian GPS spoofer to successfully create a significant timing error in a phasor measurement unit, which is a component of GPS devices responsible for estimating the magnitude and phase angle of GPS signals. Zeng et al. (2018) \cite{zeng2018all} assembled a small device from popular components with a total cost of 223 US Dollars and used it to trigger fake turn-by-turn navigation to guide victims to a wrong destination without being noticed. The authors demonstrated the attacks on real cars with 40 participants and were able to guide 38 participants to the authors' predetermined locations (95\% success rate). Zeng et al. discussed one of the limitations of their study is that it is not effective if a driver is familiar with the area. However, this may not be the case for CAVs and thus Zeng et al.'s attack model would pose a significant threat to CAVs. Recently, Regulus Cyber LTD. tested GPS spoofing on a Tesla 3 and successfully made the car's GPS display false positions on the map, and hence any attempt to find a route to a destination resulted in bad navigation \cite{Tesla3Spoof}. The total equipment cost to perform this attack was 550 US Dollars and the report also stated that ``this dangerous technology is everywhere``. Unfortunately for those who are curious, the researchers did not perform an attack when the car was on an autopilot mode. Narain et al. (2019) \cite{narain2019security} proposed an interesting approach for attackers. They first looked for data on regular patterns that exist in many cities’ road networks. Then, they used an algorithm to exploit the regular patterns and identify navigation paths that are similar to the original route (assuming that the attackers know the victim's route). Finally, the identified paths can be forced onto a target CAV through spoofed GPS signals. This attack model allows attackers to possibly bypass some defense mechanisms, such as the Inertial Navigation System (INS), which helps GPS receivers to get positions and angle updates at a quicker rate \cite{petovello2001development,el2006kalman}. The inconsistencies between the spoofed path and the original path may be negligible and the attack can be successfully executed. Also in 2019, Meng et al published an open-source GPS-spoofing generator using Software-Defined Receiver \cite{meng2019gps}. The authors claimed that their spoofing generator can cover all open-sky satellites while providing high-quality concealment, thus it can block all the legitimate signals. This would make the spoofing signals closely similar to that of the legitimate signal. Therefore, it would be difficult to detect this attack based on only the differences with surrounding GPS receivers or the signal consistency. The threat of this spoofing model to CAVs is very serious once all signals from the visible GPS satellites are spoofed \cite{meng2019gps}.

\textbf{Criteria for Defense Strategies against Attack Model 1:} Haider and Khalid (2016) \cite{haider2016survey} proposed the following criteria for effective defense strategies:
\begin{itemize}
	\item Quick Implementation: The solution can be implemented easily.
	\item Cost Effective: The solution should be affordable in either a small scale or a large scale.
	\item Prevent Simple Attack: ability to detect simple attacks.
	\item Prevent Intermediate Attacks: ability to detect intermediate type of attacks.
	\item Prevent Sophisticated Attacks: ability to detect sophisticated and advanced types of attacks, such as \cite{meng2019gps,narain2019security}.
	\item No Requirements to modify satellite transmitters: the solution does not require changes to be made on the satellite transmitters.
	\item Validation: the solution is easy to test.
	\item Interoperability: The solution works on many types of machines (we are only concerning about CAVs in this context).
\end{itemize}

\textbf{Existing Defense Strategies against Attack Model 1:} In 2003, the United States Department of Energy suggested seven simple countermeasures to detect GPS spoofing attacks \cite{warner2003gps}. The seven countermeasures are:
\begin{itemize}
	\item Monitor the absolute GPS signal strength: by recording and monitoring the average signal strength, a system may detect a GPS spoofing attack by observing that signal strengths are many orders of magnitude larger than normal signals from GPS satellites.
	\item Monitor the relative GPS signal strength: the receiver software could be programmed to record and compare signals in consecutive time frames. A large change in relative signal strength would be an indication of a spoofing attack.
	\item Monitor the signal strength of each received satellite signal: the relative and absolute signal strengths are recorded and monitored individually for each of the GPS satellites.
	\item Monitor satellite identification codes and number of satellite signals received: GPS spoofers typically transmit signals that contain tens of identification code, whereas legitimate GPS signals on the field often come from a few satellites. Keeping track of the number of satellite signals received and the satellite identification codes may help determine a spoofing attack.
	\item Check the time intervals: with most GPS spoofers, the time between signals is constant. This is not the case with real satellites. Keeping track of the time intervals between signals may be useful in detecting spoofing attack.
	\item Do a time comparison: Many GPS receivers do not have an accurate clock. By using timing data from an accurate clock to compare to the time derived from the GPS signals, we can check the veracity of the received GPS signals.
	\item Perform a sanity check: by using accelerometer and compass, a system can independently monitor and double check the position reported by the GPS receiver.
\end{itemize}

 All of the above seven countermeasures are simple and inexpensive to implement, may prevent simple attacks, do not require modification to satellite transmitters, and are inter-operable. However, they may fall short when dealing with sophisticated attacks such as \cite{meng2019gps,narain2019security}.

 Defense strategies against GPS spoofing attacks have also been studied extensively in the academic literature. Since GPS signals do not contain any information that can verify their integrity, a natural way to defend against Attack Model 1 is to use redundant information to verify the integrity of GPS signals. One example of such a method is the Receiver autonomous integrity monitoring (RAIM), a technology that uses redundant signals from multiple GPS satellites to produce several GPS position fixes and compare them \cite{brown1996receiver,hewitson2006gnss}. A RAIM system is considered available if it can receive signals from 24 or more GPS satellites \cite{van1992raim,loh1994integrity}. RAIM statistically determines whether GPS signals are faulty or malicious by using the pseudo-range measurement residual, which is the difference between the observed measurement and the expected measurement \cite{yang2016gnss}. Advanced Receiver Autonomous Integrity Monitoring (ARAIM) is a concept that extends RAIM to other constellations beyond GPS, such as GLObal NAvigation Satellite System (GLONASS), Galileo, and compass \cite{blanch2012advanced,choi2011advanced}. One criticism with ARAIM is that its availability is inconsistent if one or more satellites are not reachable \cite{qian2019impact,van1992raim}. Meng et al. (2018) proposed solutions for this problem and improved ARAIM availability up to 98.75\% \cite{qian2019impact,meng2019improved}. There are many other validation mechanisms, which all make use of additional satellites or side-channel information. For example, O’Hanlon et al. described how to estimate the expected GPS signal strength and compared against the observed signal strength to validate GPS signals \cite{o2013real}. Furthermore, a defense system can monitor GPS signals to ensure that the rate of change is within a threshold. Montgomery proposed a defense approach that uses a dual antenna receiver that employs a receiver-autonomous angle-of-arrival spoofing countermeasure \cite{montgomery2011receiver}. The main idea is to measure the difference in the signal's phase between multiple antennas referenced to a common oscillator. Other examples of countermeasures can be found in the survey paper by Haider and Khalid (2016) \cite{haider2016survey}.

\textbf{Challenges for defending against Attack Model 1:} Even though several countermeasures have been proposed in the literature, their effectiveness against newer attack strategies, such as \cite{meng2019gps,narain2019security} (2019), is unknown. The attack strategy in \cite{meng2019gps} is especially dangerous for CAVs. Therefore, finding effective countermeasures for these 2019-born attack models is a current research challenge.

\textbf{Attack Model 2 - GPS jamming:} Since radio signals from the satellites are generally weak, jamming can be achieved by firing strong signals that overwhelm GPS receiver, so that the legitimate signals can not be detected \cite{hu2009study}. Examples that highlight the risks of GPS jamming have been reported. In 2013, a New Jersey man was arrested for using a \$100 GPS jamming device plugged into the cigarette lighter in his company truck \cite{helfrick2014question}. The man's motive was to jam his company truck's GPS signal to hide from his employer. However, the device was reported to be powerful enough to interfere with GPS signals at the nearby Newark airport. Even though GPS jamming devices are illegal for civilian use, they can easily be found on online retailers such as eBay \cite{coffed2014threat}. GPS jamming is less dangerous than GPS spoofing in the sense that attackers have higher control over GPS receivers with a spoofing attack. However, GPS jamming attack can cause disruptions of service and is essentially a Denial-of-Service attack.

\textbf{Criteria for Defense Strategies against Attack Model 2:} Based on our understanding of the attack model and study on the defense techniques, we suggest the following criteria.
\begin{itemize}
	\item Attack Detection: The solution should be able to detect a jamming attack in a timely manner to ensure the safety of CAVs.
	\item Signal Filter: The solution should be able to filter out the attack signals so that the vehicle can still operate under certain attack scenarios and avoid disruption of service.
\end{itemize}

\textbf{Existing Defense Strategies against Attack Model 2:} Many GPS receiver modules have implemented anti-jamming measures that target unintentional interference from every-day electronic devices. However, Hunkeler et al. (2012) \cite{hunkeler2012effectiveness} have shown that these countermeasures are ineffective against intentional attacks. For example, the NEO-6 GPS receiver has an integrated anti-jamming module that provides data to assess the likelihood that a jamming attack is ongoing. Hunkeler et al. showed that the parasitic signal from the GPS jammer interfered with the NEO-6 receiver in such a way that the receiver could not function while the anti-jamming module reported very low probability for a jamming attack. Several studies have demonstrated how calculate the probability of intentional GPS jamming attack just by using information from GPS receivers \cite{hunkeler2012effectiveness,zhang2012anti,sun2005self}. Unfortunately, detection of GPS jamming does not prevent disruption of service, which is the main objective of a GPS jamming attack. L3Harris Technologies, Inc. developed a technology, Excelis Sentry 1000, that can detect sources of interference to support timely and effective actionable intelligence \cite{pattinson2017standardisation}. The Excelis Sentry 1000 systems can be strategically placed around high-risk areas to instantaneously sense and triangulate the location of jamming sources \cite{coffed2014threat}. However, this defense strategy may not be effective for CAVs, whose operating location is not predictable.

Mukhopadhyay (2007) \cite{mukhopadhyay2007augmentation} and Purwar et al. (2016) \cite{purwar2016gps} proposed methods to reduce the jamming signals and estimate the legitimate GPS signals. Mukhopadhyay used a the Adaptive Array Antenna technology and the Least Mean Squared (LMS) algorithm to maximize the chance of collecting the desired signals and the chance of rejecting jamming signals. The Adaptive Array Antenna technology are antenna arrays that have integrated signal processing algorithms that can identify spatial signal signatures such as the direction of arrival (DOA) of the signal, and use them to calculate beamforming vectors in order to track and locate the antenna beam. Mukhopadhyay used the LMS algorithm, which is a member of a family of stochastic gradient algorithms, to further accurately calculate the DOA. The DOA information would then contribute to determining the signals being rejected or accepted. Purwar et al. (2016) \cite{purwar2016gps} proposed the Turbo Coding method for counter jamming. In Turbo Coding, The Turbo encoder at the GPS satellites that send the original GPS data is encoded, then modulated, and passed over a noisy channel. Then, the encoded data arrives at the GPS receivers along with noise and jamming signals. The receiver demodulates the distorted GPS signals and then the Turbo Decoder decodes demodulated signals to retrieve the original GPS signals. 
This technique has two major weaknesses. First, the results in \cite{purwar2016gps} demonstrated that as the strength of jamming signals increases, their method becomes less effective. Specifically, when the jamming signals is about 14 times stronger than the legitimate signals, the method cannot recover the legitimate signals. Second, this method requires modification to the GPS satellites.

\textbf{Challenges for defending against Attack Model 2:} The state-of-the-art countermeasure was published by Purwar et al. (2016) \cite{purwar2016gps}. It has two major weaknesses that would be problematic to be implemented for CAVs. The method requires modification to the GPS satellites and is only effective if the jamming signals are not too strong.

\subsection{Attacks on cameras}

\textbf{Attack Model 1 - Camera blinding:} On CAVs, cameras commonly provide inputs for deep learning models for the task of object detection. Attackers aim for a denial of this service by blinding a camera with extra light. Petit et al. (2014) \cite{petit2015remote} experimented with a blinding attack on a MobilEye C2-270 camera installed on a non-automated car's windshield. The researchers showed that a quick burst of 650 nm laser was able to almost fully blind the camera and the camera never recovered from the blindness. The 940 nm 5x5 LED matrix and 850 nm LED spot also achieved the same result, but the camera was able to recover after more than 5 seconds. It is important to note that the MobilEye C2-270 camera is not used for full vehicle automation (SAE Level 5) but function-specific automation (SAE Level 1 to 3). Yan et al. (2016) \cite{yan2016can} experimented with similar attacks and was also able to blind a camera permanently (the authors did not specify the camera). This paper also pointed out that LED and Laser beam could blind a camera but Infrared LED could not because of the narrow frequency band filters. We have not found any other study on this attack model.

\textbf{Criteria for Defense Strategies against Attack Model 1:} Based on our understanding of the attack model and study on the defense techniques, we suggest the following criteria.
\begin{itemize}
	\item Low cost: The solution should not require expensive modification to the vehicle
	\item Generalization: The solution should work on as many attack wavelengths as possible. Petit et al. and Yan et al. showed that this attack model is possible for laser and LED with different wavelengths. A solution that works for all attack wavelengths would be ideal.
\end{itemize}

\textbf{Existing Defense Strategies against Attack Model 1:} Petit et al. \cite{petit2015remote} also suggested two countermeasures in their study. The first countermeasure is to use redundancy by installing multiple cameras that have overlapping coverage. This is effective because laser and LED spot have small beam width, making it difficult to attack multiple cameras spontaneously. This defense strategy cannot mitigate the risk of Attack Model 1 completely, but it makes attackers spend more effort on a successful attack. Nevertheless, the cost to implement this solution grows in direct proportion to the number of extra cameras. The second countermeasure is to integrate a removable near-infrared-cut filter into a camera. This is a technology that is available on security cameras and can filter near-infrared light on request. This solution can potentially satisfy both criteria, but would need implementation and experiments to be verified. In 2020, DH and Ansari proposed a detection method by using predictive analytics to predict the future next frames captured by the cameras and then compare the received frames with the predicted frames \cite{dh2020autonomous}. DH and Ansari's solution has the potential to satisfy both the criteria we mentioned.

\textbf{Challenges for defending against Attack Model 1:} Petit et al.'s suggested countermeasures are the only ones that we found in the literature. Yan et al, (2016) \cite{yan2016can} did not discuss any countermeasure in their experiment. Petit et al.'s proposed strategy of using near-infrared-cut filter into cameras has high potential, but need further experiments and validations. Similarly, DH and Ansari's solution has the potential to be a generalized solution and may be implemented easily, but need further validations due to its newness.

\textbf{Attack Model 2 - Adversarial Images:} An adversary may carefully make small perturbations on the images that cameras observe and cause the artificial-intelligence algorithms (used for CAV's vision) to generate incorrect predictions. Even though the ultimate targets of this type of attack are the deep learning models that reside in ECUs, cameras are convenient channels for attackers to inject adversarial images. In 2017, Google researchers were able to create stickers called adversarial patch \cite{brown2017adversarial} with patterns that can deceive artificial intelligence algorithms. These stickers may be printed out and attached to important transportation objects, such as road signs. Lu et al. (2017) \cite{lu2017no} experimented such an attack by taking 180 photos of compromised stop signs with an iPhone 7 from a moving vehicle. They found that a trained neural network classified most of the pictures correctly, which meant that the attack model was not effective. In contrast, Eykholt et al. (2018) \cite{eykholt2018robust} were a lot more successful with their experiments. They experimented with a camera in a lab setting and a camera on a moving vehicle (non-CAV). By decorating stop signs with small black-and-white stickers, the authors made a state-of-the-art algorithm fail to recognize the stop signs 100\% of the time in a lab setting and 84.8\% on a moving vehicle. Li and Gerdes (2019) \cite{kelarestaghi2019intelligent} experimented with the same type of attack by using electromagnetic interference in a remote manner and without physical modification to the stop signs, which makes the attack easier to launch and harder to detect. We have not found any experiment on CAVs, but this attack model, as demonstrated in a moving vehicle setting \cite{eykholt2018robust}, should also apply to CAVs. There are several methods that attackers can use to know how to tamper with objects such as road signs. These methods are called universal adversarial perturbations and are described in \cite{moosavi2017universal,goodfellow2014explaining,kos2018adversarial,sitawarin2018darts}. Recently, an IBM research group released an open-source Python library, called Adversarial Robustness 360 Toolbox, for generating and defending against adversarial images \cite{art2018}. This attack model is especially dangerous because it may make CAVs ignore alerts or read the wrong speed limits from road signs.

\textbf{Criteria for Defense Strategies against Attack Model 2:} Based on our understanding of the attack model and study on the defense techniques, we suggest the following criteria.
\begin{itemize}
	\item Low cost and easy implementation: The solution should not require expensive modification to the vehicle
	\item Generalization: The solution should work on many types of image perturbations.
	\item Computationally efficient: The solution should be calculated efficiently to serve real-time object-detection purposes.
\end{itemize}

\textbf{Existing Defense Strategies against Attack Model 2} Securing Machine Learning models against adversarial images have been discussed extensively. This can be achieved by several techniques such as pre-processing inputs \cite{zantedeschi2017efficient,xu2017feature,dziugaite2016study,guo2017countering}, adding adversarial samples to training data \cite{szegedy2013intriguing,miyato2015distributional,jiang2020poisoning}, and utilizing run-time information to detect abnormal inputs \cite{buckman2018thermometer}. All of these methods are algorithm-based and can probably be integrated into the codes on the ECUs that handle object-detection from camera data. To our knowledge, no research has studied the extent of generalization of these algorithms, as well as their theoretical and practical computational complexity. Such a comparison study for these defense techniques should be performed for further discussions on this topic.

\textbf{Challenges for defending against Attack Model 2:} The attack and defense of Attack Model 2 have been studied extensively in terms of methodologies and have been tested with general images. A comparison study for these defense techniques in terms of generalization and computational efficiency should be performed to server further discussions on this topic.

\subsection{Attacks on communication mechanisms}

\textbf{Attack Model 1 - Falsified information:} Attackers aim to send falsified information through the V2V and V2I communications to disrupt CAVs' operation and traffic flow. This can be achieved by impersonation or Sybil attacks. In an impersonation attack, attackers steal the identity of legitimate CAVs and broadcast falsified information. This could be achieved if the connection protocol lacks a strong authentication method. Chim et al. (2009) \cite{chim2009security} described in detail one way to impersonate another CAV's identity in section IV of their paper. In a Sybil attack, attackers create a large number of identities and use them to send falsified information over a network and make the falsified information appear to be popular and legitimate. For example, Rawat et al. (2019)  \cite{rawat2012vanet} described a scenario in a VANET network, where fake identities are made to look like they surround a target vehicle, making the target vehicle think that there is a traffic jam.

\textbf{Criteria for Defense Strategies against Attack Model 1:} This attack model can be mitigated by implementing strong authentication methods that may be specified in the V2V and V2I network protocols. Based on our understanding of the attack model and study on the defense techniques, we suggest the following criteria.

\begin{itemize}
	\item Low cost and easy implementation: Solutions that are easier to implement and cost less are preferable.
	\item Computationally efficient: The solution should be calculated efficiently to serve real-time authentication.
\end{itemize}

\textbf{Existing Defense Strategies against Attack Model 1:} Several authentication methods have been proposed, such as \cite{grover2011sybil,whyte2013security,alimohammadi2015sybil}. Grover et al. (2011) \cite{grover2011sybil} proposed an authentication scheme that uses neighboring vehicles in VANET.  There are four phases:
\begin{itemize}
	\item Periodic Communication: each vehicle on the road periodically broadcasts and receives beacon packets. This phase is to announce a vehicle's presence to all vehicles on the road.
	\item Group construction of neighboring nodes: when a vehicle collects enough beacon messages from other vehicles, it makes a record of neighboring nodes in the form of groups at regular intervals of time.
	\item Exchange groups with other nodes in vicinity: after significant duration of time, these vehicles exchange their neighboring nodes record with each other in vicinity.
	\item Identify the vehicles comprising similar neighboring nodes: after receiving the records from other vehicles, each vehicles may detect malicious vehicles by observing that the abnormal vehicles exist in neighboring nodes for duration greater than a threshold.
\end{itemize}
Grover's approach may satisfy both of aforementioned criteria, but may fall short when there are not enough vehicles on the road to exchange information. Whyte et al. (2013) \cite{whyte2013security} proposed the Security Credential Management System (SCMS) that implements a public-key infrastructure (PKI) with some features for providing privacy. The PKI is used to authenticate vehicles through V2I connections. Two major drawbacks of this approach are that a PKIs are expensive to implement, and that V2I communication may add significant delays to the authentication process. Alimohammadi et al. (2015) \cite{alimohammadi2015sybil} proposed a secure protocol based on a light weight group signature scheme. For a short time and secure group communication, the Boneh-Shacham algorithm is used for short group signature scheme and batch verification. Hubaux (2004) \cite{hubaux2004security} suggested that all vehicles use electronic license plates to allow the wireless authentication of CAVs. Zhang et al. (2008) \cite{zhang2008efficient} proposed an authentication scheme that would efficiently authenticate CAVs by using some roadside infrastructures. However, Chim et al. (2009) \cite{chim2009security} claimed that their attack model would penetrate Zhang et al.'s method. In 2020, Zhao et al. \cite{zhao2020efficient} proposed a protection mechanism that consists of an offline phase and an online phase. The offline phase establishes matrices and parameters to support attack detection and decision making during the online phase. All the methods that require PKI and roadside infrastructures are costly and difficult to implement.

\textbf{Challenges for defending against Attack Model 1:} Strong authentication methods that are based on public-key encryption would require a public key infrastructure. However, as van der Heijden et al. (2018) discussed \cite{van2018survey}, PKIs suffer from problems such as being expensive to implement in a large region and may not provide real-time authentication. Furthermore, Chim et al. \cite{chim2009security} pointed out that the processing units on CAVs may not be able to process the authenticating messages in real time.

\textbf{Attack Model 2 - Denial of Service:} Attackers perform Denial-of-Service (DoS) attack and Distributed Denial-of-Service (DDoS) attack on the communication mechanisms. Several studies have demonstrated how VANETS and V2I networks are vulnerable to DoS and DDoS attacks\cite{pathre2013identification,pathre2013novel,douligeris2004ddos,hasbullah2010denial}. All of these attacks are intended to confuse CAVs' operations and disrupt traffic flows. DDoS attacks on V2I networks are likely to disrupt transportation in a large region, especially if the infrastructure provides critical information to control traffic flows. Ekedebe et al. (2015) \cite{Ekedebe2015simulation} experimented with DoS attacks on V2I network by using a simulated environment and showed that DoS attacks can prevent all messages from being sent to vehicles in the network. V2V networks are also possible targets for DoS and DDoS attacks because they have limited connection bandwidth. For example, the DSRC standard specifies that a node must wait to transmit signals until the DSRC channel is idle \cite{blum2006fast}. To exploit this limitation, attackers may constantly transmit noises through the DSRC channel to keep it always busy and thus to prevent legitimate signals from being delivered. Leinimuller et al. (2006) \cite{leinmuller2006position} described another variation of DoS attacks on V2V networks, where an adversarial vehicle in a VANET network can falsify its position information to intercept message packets between other vehicles in the network.

\textbf{Existing Defense Strategies against Attack Model 2:} Defending against DoS and DDoS attacks on V2I and V2V network requires a secured V2I or V2V network architecture and protocols. Just like DoS attacks on V2I networks resemble those on a traditional centralized network, defending strategies for a traditional centralized network can be applied to defend V2I networks. The defense strategies for a general centralized network can be found in Douligeris's survey paper (2004) \cite{douligeris2004ddos}, which include intrusion prevention, intrusion detection, intrusion response, and intrusion tolerance techniques. Singh et al. (2018) \cite{singh2018ml} discussed a machine-learning based approach to detect DDoS attacks on V2I networks. Besides this study, We have not found any other study that is specific for V2I networks. To secure V2V networks, several papers have proposed secured architecture. For example, Blum et al. (2006) \cite{blum2006fast} proposed the CARAVAN architecture, Plossl (2016) \cite{plossl2006towards} proposed an architecture that uses a certified GALILEO receiver to achieve reliable time and position information, Hubaux (2004) \cite{hubaux2004security} suggested using electronic license plates to authenticate CAVs.

\textbf{Challenges for defending against Attack Model 2:} All the aforementioned studies are based on theoretical frameworks and have not been tested in a realistic CAV environment. An experimental study, even if performed on simulated but realistic data, would be valuable to determine suitable methods to defend against DoS and DDoS attacks on V2V and V2I networks.

\section{Research Challenges, Trends, and Open Issues}
In this section, we first highlight the challenges and research trends that we have observed and discussed in sections II and III. These observations all came from academic research. Next, we present official publications and reports from the industry that are related to the cybersecurity of CAVs. We searched for publications and reports from corporations that are involved in the developments of CAVs.

\subsection{Academic Research}
\textbf{Research trends:} We have observed the following research trends happening on the topic of security of CAVs.
\begin{itemize}
    \item The trend in developing new attack models largely follows the remote attack pattern by targeting sensors, cameras, and communication mechanisms. We have observed that most of the recently developed attacks can be performed from a remote distance, either from the roadside or from other vehicles. Examples of such attack can be found in \cite{petit2015remote,narain2019security,meng2019gps,pathre2013identification}.
    \item The trend in developing new defense strategies is to utilize machine learning and anomaly-based intrusion detection systems. The requirements for new defense strategies often include the capability of real-time response, low expense, and less modification to CAVs' architecture.
    \item There is an obvious need to develop secured mechanisms to update the software on ECUs. All the current updating mechanisms are easily abused by attackers to compromise CAVs, such as those attack models mentioned in \cite{nilsson2008creating,mulliner2012read}.
    \item A decentralized public key infrastructure (PKI) will be necessary to improve the security of V2V and V2I communications \cite{van2018survey,techxplore}. While operating, a CAV may meet many other CAVs in a short time and may need to exchange information through V2V communications. To mitigate the attack model 1 discussed in section III-H, authentication based on public-key encryption is necessary. To achieve this, a CAV will need to request the public key of other CAVS from a PKI. Decentralizing PKI is necessary to reduce the latency before the vehicle receives the necessary keys to achieve real-time authentication.
\end{itemize}

\textbf{Research Challenges:} From the research challenges discussed in section III, we organize three categories for the open security issues on CAVs. The specific challenges and their related publications are presented in Table III.
\begin{itemize}
    \item \textbf{Unsolved attack models:} there are several attack models in the literature that we have not found a corresponding defense strategy in the literature. Some of these attack models pose significant and realistic threats to CAVs, such as the GPS Spoofing methods discussed in\cite{meng2019gps} and \cite{narain2019security}, both published in 2019. These challenges are presented in the first column of Table III.
    \item \textbf{Attack models needing further experiments}: there are recently developed attack models that we are unable to assess the level of threat and effectiveness in a realistic environment of CAVs. Further experiments under a realistic environment will provide the community with a better understanding of the impact of these attack models and incentivize research for defense strategies. These challenges are presented in the second column of Table III.
    \item \textbf{Defense strategies needing further experiments}: there are many defense strategies proposed only theoretically or have only been tested under unrealistically simulated environments. Further experiments under realistic environment will help validate these strategies and identify the most suitable ones. These challenges are presented in the third column of Table III.
\end{itemize}

\begin{table*}[h]
 \caption{}
  \begin{tabular}{|>{\raggedright\arraybackslash}p{4.5cm}|p{5.5cm}|p{6.0cm}|}
\hline
      \multicolumn{1}{|>{\centering\arraybackslash}m{4.5cm}|}{\cellcolor{Gray}\textbf{Unsolved attack models}}
    & \multicolumn{1}{>{\centering\arraybackslash}m{5.5cm}|}{\cellcolor{Gray} \textbf{Attack models needing further experiments}}
    & \multicolumn{1}{>{\centering\arraybackslash}m{6.0cm}|}{\cellcolor{Gray} \textbf{Defense strategies needing further experiments}}\\
  \hline
  \begin{itemize}[leftmargin=*]
  \item No study have proposed a defense layer to secure the OBD port that aims to distinguish legitimate from malicious OBD devices.
  
  \item Recently developed GPS Spoofing techniques \cite{meng2019gps,narain2019security} still lack an effective countermeasure
  
  \item Several efficient methods to detect GPS jamming are described in \cite{hunkeler2012effectiveness,zhang2012anti,sun2005self}, but none of them could filter out the jamming signals to obtain the legitimate signals. Maintaining GPS service under GPS jamming attacks remains a challenge.
  
  \item A strong and efficient authentication method is necessary to defend V2V and V2I networks from Sybil and impersonation attacks. Public-key encryption can provide authenticity, but is difficult to implement on V2V and V2I due to the reasons presented in \cite{van2018survey}. Further studies are needed to find an adequate solution that overcomes all the issues discussed in \cite{van2018survey}.
\end{itemize}
  & 
  \begin{itemize}[leftmargin=*]
  \item The LiDAR jamming attack model discussed in \cite{stottelaar2015practical} has not been experimented in a CAV environment.
  \item Yan et al. \cite{yan2016can} failed to apply the DRFM technique to inject counterfeit signals to a radar sensor on a Tesla Model S. Further experiments are needed to determine whether this is also the case for other CAVs and whether the challenges in this attack model can be overcome.
  \item Radar jamming attacks have been widely studied for manned and unmanned aerial vehicles \cite{buehler2014airborne,stott1994digital,Lothes1990jammingbook}, but have not been experimented on CAVs.
  \item Although the experiments on GPS jamming have been demonstrated in \cite{hu2009study,helfrick2014question} and the jamming devices can be obtained with relative ease, the potential and threat level of such an attack on CAVs have not been studied.
\end{itemize}
  &
  \begin{itemize}[leftmargin=*]
  \item Many defense strategies have been proposed for defending against CAN attack through the OBD port, but most of them receive criticisms \cite{shin2017fingerprinting,cho2016fingerprinting,gmiden2016intrusion,tyree2018exploiting,tomlinson2018towards}. A comparison study is needed in this topic.
  
  \item Several IDS-based algorithms have been proposed to detect abnormal CAN network packets that are sent from compromised ECUs \cite{miller2014survey,muter2011entropy,tyree2018exploiting}. However, it is still unclear how effective these approaches are under a realistic CAV environment.
  
  \item Two frameworks are proposed in \cite{seshadri2006scuba} and \cite{nilsson2008framework} that aim to secure firmware updates over the air (FOTA). They still need further testings and experiments.
  
  \item Recently proposed defense strategies against LiDAR Spoofing  (\cite{matsumura2018secure} and \cite{davidson2016controlling}) need further experiments.
  
  \item Camera blinding attacks have been successfully demonstrated on CAVS \cite{petit2015remote,yan2016can}. Petit et al.’s countermeasures \cite{petit2015remote} are the only current countermeasures against this attack model. Further studies to validate this defense strategy is needed.
  
  \item Adversarial image attacks, as experimented in \cite{eykholt2018robust,kelarestaghi2019intelligent}, have many proposed countermeasures in the literature, which can be found in \cite{art2018}. Implementation of these countermeasures need to be validated.
  
  \item There exists countermeasures against DDoS attacks on V2V networks \cite{blum2006fast,plossl2006towards,hubaux2004security}. However, they are all based on theoretical frameworks and need validation in a realistic CAV environment.
  
\end{itemize}
  \\ 
  \hline
  
\end{tabular}
\end{table*}

\subsection{Industrial development}
In July 2019, a coalition of 11 companies that are involved in the development of CAVs published a whitepaper titled ``Safety First For Automated Driving``~\cite{safetyfirst2019}. The 11 companies include Audi, Intel, Volkswagen, Baidu, BMW, Aptiv, Fiat Chrysler Automobiles, Daimler, Continental, Infineon, and HERE. The whitepaper describes a thorough approach to make CAVs operate safer, of which security is a subtopic. With this publication, the authors aimed to build a guideline for the development of safer and more secure CAVs. The authors claimed to continuously update this whitepaper by including detailed solutions for defined problems in the future. They hope that the whitepaper will become an international standard for the development of CAVs. Regarding security issues, the whitepaper~\cite{safetyfirst2019} recommends two approaches:
\begin{itemize}
    \item Using Secure Development Lifecycle (SDL), which is a process for integrating security into the product development and product maintenance processes. SDL practices are divided into preliminaries, development, and sustainment. Regarding preliminaries, a CAV development team is required to be sufficiently trained with knowledge of security issues, policies, procedures, and guidelines. Development includes security practices in software engineering, such as code review, penetration testing, and threat modeling. Sustainment practices ensure that CAVs continue to operate safely after release by having an effective incident response system and continuous updates.
    \item Regarding the machine learning models embedded in CAVs, the authors conducted a brief survey on the challenges of building safety-ensured machine learning models and suggested many guidelines. This section is presented in Appendix B of the whitepaper~\cite{safetyfirst2019}. In short, the guidelines involve the processes of selecting data for model training and testing, architecture design of models, model evaluation, and deployment and monitoring. Interestingly, we think that building defense mechanisms against adversarial image attacks (section 3E, attack model 2) fits into the paradigm of data selection and model validation that the whitepaper describes.
\end{itemize}

Another coalition of Ford, Lyft, Uber, Volvo, and Waymo, named the Self-Driving Coalition for Safer Streets also presented several publications related to CAV's security on their website \cite{selfdrivingcoalition}. Most notable is Waymo's safety report titled ``On the Road to Fully Self-Driving`` \cite{waymo2017report}. The report claimed that Waymo applied approached such as building redundant security measures for critical systems and limiting communication between critical systems. However, the specific implementation of these approaches was not described. We have not found any related safety report from other members of this coalition.

Regarding secure communication for CAN Bus, Guardknox Cyber Technologies Ltd. developed a patented security approach, named Communication Lockdown, to enforce a formally verified and deterministic configuration of communication among the CAN Bus \cite{GuardKnox}. Guardknox Cyber Technologies Ltd. claims that this technology can provide zero false positives with minimal integration and no vehicular hardware modification.

From our observations, research and implementation of security issues are in the early stage of development among the CAV corporations. We have not found any strong connection between academic research and the industry's implementation of CAV-related security issues. There is no evidence to show that CAVs on the market have been updated to defend against the novel attack models that the research community has found.

\section{Conclusion}
Modern innovations of Connected and Autonomous Vehicles (CAVs) are transforming transportation and gaining a lot of public attention. While CAVs have enormous potentials to change human life, they pose significant security concerns and are vulnerable targets for attackers. Therefore, interest in the security of CAVs has been increasing rapidly. During the last decade, many attack models and defense strategies for CAVs have been discussed and experimented with. In this paper, we studied 184 papers from 2000 to 2019 to understand state-of-the-art security issues with CAVs. CAVs are prone to attacks on many of their components. These attacks can render CAVs out of service or give attackers control over CAVs. Some attack models are published recently and are seriously threatening to CAVs. We have presented readers with a comprehensive review of the security challenges of CAVs and corresponding state-of-the-art countermeasures. Furthermore, we organized the attack models based on their target components, access requirements, and attack motives. Finally, we have identified some research challenges and future directions that researchers can contribute to so that CAVs become secure and trustworthy to the general public.

\pagebreak
\bibliographystyle{ieeetr}
\bibliography{reference.bib}

\begin{thebibliography}{100}

\bibitem{figueiredo2001towards}
L.~Figueiredo, I.~Jesus, J.~T. Machado, J.~R. Ferreira, and J.~M. De~Carvalho,
  ``Towards the development of intelligent transportation systems,'' in {\em
  IEEE Intelligent Transportation Systems (ITSC) (Cat. No. 01TH8585)},
  pp.~1206--1211, IEEE, 2001.

\bibitem{becker2000sensor}
J.~C. Becker and A.~Simon, ``Sensor and navigation data fusion for an
  autonomous vehicle,'' in {\em Proceedings of the IEEE Intelligent Vehicles
  Symposium (Cat. No. 00TH8511)}, pp.~156--161, IEEE, 2000.

\bibitem{bansal2017forecasting}
P.~Bansal and K.~M. Kockelman, ``Forecasting americans’ long-term adoption of
  connected and autonomous vehicle technologies,'' {\em Transportation Research
  Part A: Policy and Practice}, vol.~95, pp.~49--63, 2017.

\bibitem{uhlemann2015autonomous}
E.~Uhlemann, ``Autonomous vehicles are connecting...[connected vehicles],''
  {\em IEEE Vehicular Technology Magazine}, vol.~10, no.~2, pp.~22--25, 2015.

\bibitem{williams1988prometheus}
M.~Williams, ``Prometheus-the european research programme for optimising the
  road transport system in europe,'' in {\em IEE Colloquium on Driver
  Information}, pp.~1--1, IET, 1988.

\bibitem{benmimoun2009demonstration}
A.~Benmimoun, M.~Lowson, A.~Marques, G.~Giustiniani, and M.~Parent,
  ``Demonstration of advanced transport applications in citymobil project,''
  {\em Transportation Research Record}, vol.~2110, no.~1, pp.~9--17, 2009.

\bibitem{suzuki2010development}
Y.~Suzuki, T.~Hori, T.~Kitazumi, K.~Aoki, T.~Fukao, and T.~Sugimachi,
  ``Development of automated platooning system based on heavy duty trucks,'' in
  {\em 7th IFAC Symposium on Advances in Automotive Control}, 2010.

\bibitem{van2012cooperative}
E.~van Nunen, M.~R. Kwakkernaat, J.~Ploeg, and B.~D. Netten, ``Cooperative
  competition for future mobility,'' {\em IEEE Transactions on Intelligent
  Transportation Systems}, vol.~13, no.~3, pp.~1018--1025, 2012.

\bibitem{sae2018taxonomy}
{SAE international}, ``{Taxonomy and definitions for terms related to driving
  automation systems for on-road motor vehicles}.''
  \url{https://www.sae.org/standards/content/j3016\_201806/}, 2018.

\bibitem{wyglinski2013security}
A.~M. Wyglinski, X.~Huang, T.~Padir, L.~Lai, T.~R. Eisenbarth, and
  K.~Venkatasubramanian, ``Security of autonomous systems employing embedded
  computing and sensors,'' {\em IEEE micro}, vol.~33, no.~1, pp.~80--86, 2013.

\bibitem{ieee2010ieee}
{IEEE Standards Association}, ``{IEEE Standard for Information Technology-Local
  and Metropolitan Area Networks-Specific Requirements-part 11: Wireless LAN
  Medium Access Control (MAC) and Physical Layer (PHY) Specifications Amendment
  6: Wireless Access in Vehicular Environments},'' {\em {IEEE Std 802.11-2016
  (Revision of IEEE Std 802.11-2012)}}, 2010.

\bibitem{durrant2006simultaneous}
H.~Durrant-Whyte and T.~Bailey, ``{Simultaneous localization and mapping: part
  I},'' {\em IEEE robotics \& automation magazine}, vol.~13, no.~2,
  pp.~99--110, 2006.

\bibitem{bailey2006simultaneous}
T.~Bailey and H.~Durrant-Whyte, ``{Simultaneous localization and mapping
  (SLAM): Part II},'' {\em IEEE robotics \& automation magazine}, vol.~13,
  no.~3, pp.~108--117, 2006.

\bibitem{pilipovic2014toward}
M.~Pilipovic, D.~Spasojevic, I.~Velikic, and N.~Teslic, ``Toward intelligent
  driver-assist technologies and piloted driving: Overview, motivation and
  challenges,'' in {\em Proceedings of the X International Symposium on
  Industrial Electronics (INDEL’14)}, 2014.

\bibitem{bohm2008supporting}
A.~Bohm and M.~Jonsson, ``Supporting real-time data traffic in safety-critical
  vehicle-to-infrastructure communication,'' in {\em 33rd IEEE Conference on
  Local Computer Networks (LCN)}, pp.~614--621, IEEE, 2008.

\bibitem{xu2004vehicle}
Q.~Xu, T.~Mak, J.~Ko, and R.~Sengupta, ``{Vehicle-to-vehicle safety messaging
  in DSRC},'' in {\em Proceedings of the 1st ACM international workshop on
  Vehicular ad hoc networks}, pp.~19--28, ACM, 2004.

\bibitem{biswas2006vehicle}
S.~Biswas, R.~Tatchikou, and F.~Dion, ``Vehicle-to-vehicle wireless
  communication protocols for enhancing highway traffic safety,'' {\em IEEE
  communications magazine}, vol.~44, no.~1, pp.~74--82, 2006.

\bibitem{miller2014survey}
C.~Miller and C.~Valasek, ``A survey of remote automotive attack surfaces,'' in
  {\em {Black Hat USA}}, p.~94, 2014.

\bibitem{thing2016autonomous}
V.~L. Thing and J.~Wu, ``Autonomous vehicle security: A taxonomy of attacks and
  defences,'' in {\em IEEE International Conference on Internet of Things
  (iThings) and IEEE Green Computing and Communications (GreenCom) and IEEE
  Cyber, Physical and Social Computing (CPSCom) and IEEE Smart Data
  (SmartData)}, pp.~164--170, IEEE, 2016.

\bibitem{haider2016survey}
Z.~Haider and S.~Khalid, ``{Survey on effective GPS spoofing
  countermeasures},'' in {\em Sixth International Conference on Innovative
  Computing Technology (INTECH)}, pp.~573--577, IEEE, 2016.

\bibitem{parkinson2017cyber}
S.~Parkinson, P.~Ward, K.~Wilson, and J.~Miller, ``Cyber threats facing
  autonomous and connected vehicles: Future challenges,'' {\em {IEEE
  transactions on Intelligent Transportation Systems}}, vol.~18, no.~11,
  pp.~2898--2915, 2017.

\bibitem{tomlinson2018towards}
A.~Tomlinson, J.~Bryans, and S.~A. Shaikh, ``Towards viable intrusion detection
  methods for the automotive controller area network,'' in {\em 2nd ACM
  Computer Science in Cars Symposium}, 2018.

\bibitem{van2018survey}
R.~W. van~der Heijden, S.~Dietzel, T.~Leinm{\"u}ller, and F.~Kargl, ``Survey on
  misbehavior detection in cooperative intelligent transportation systems,''
  {\em IEEE Communications Surveys \& Tutorials}, vol.~21, no.~1, pp.~779--811,
  2018.

\bibitem{gopalakrishna2015connected}
D.~Gopalakrishna, V.~Garcia, A.~Ragan, T.~English, S.~Zumpf, R.~Young,
  M.~Ahmed, F.~Kitchener, N.~U. Serulle, E.~Hsu, {\em et~al.}, ``{Connected
  vehicle pilot deployment program phase 1, concept of operations (ConOps),
  ICF/Wyoming.},'' tech. rep., {United States Department of Transportation,
  Intelligent Transportation Systems Joint Program Office}, 2015.

\bibitem{kitchener2017connected}
F.~Kitchener, T.~English, D.~Gopalakrishna, V.~Garcia, A.~Ragan, R.~Young,
  M.~Ahmed, D.~Stephens, N.~U. Serulle, {\em et~al.}, ``Connected vehicle pilot
  deployment program phase 2, data management plan-wyoming,'' tech. rep.,
  United States Department of Transportation, Intelligent Transportation
  Systems Joint Program Office, 2017.

\bibitem{wyoming}
{Wyoming Department of Transportation}, ``{Wyoming DOT Connected Vehicle
  Pilot}.'' \url{https://wydotcvp.wyoroad.info/}.

\bibitem{NYC}
{New York City Department of Transportation}, ``{NYC Connected Vehicle Project
  For Safer Transportation}.'' \url{https://cvp.nyc/}.

\bibitem{Tampa}
{Tampa Hillsborough Expressway Authority}, ``{THEA Connected Vehicle Pilot}.''
  \url{https://www.tampacvpilot.com/}.

\bibitem{UK}
{Centre for Connected and Autonomous Vehicles}, ``{UK Connected \& Autonomous
  Vehicle Research \& Development Projects 2018}.''
  \url{https://assets.publishing.service.gov.uk/government/uploads/system/uploads/attachment\_data/file/737778/ccav-research-and-development-projects.pdf}.

\bibitem{west2016moving}
D.~M. West, ``Moving forward: Self-driving vehicles in china, europe, japan,
  korea, and the united states,'' {\em Center for Technology Innovation at
  Brookings}, September 2016.

\bibitem{jo2015development}
K.~Jo, J.~Kim, D.~Kim, C.~Jang, and M.~Sunwoo, ``{Development of autonomous
  car—Part II: A case study on the implementation of an autonomous driving
  system based on distributed architecture},'' {\em IEEE Transactions on
  Industrial Electronics}, vol.~62, no.~8, pp.~5119--5132, 2015.

\bibitem{petit2014potential}
J.~Petit and S.~E. Shladover, ``Potential cyberattacks on automated vehicles,''
  {\em IEEE Transactions on Intelligent Transportation Systems}, vol.~16,
  no.~2, pp.~546--556, 2014.

\bibitem{checkoway2011comprehensive}
S.~Checkoway, D.~McCoy, B.~Kantor, D.~Anderson, H.~Shacham, S.~Savage,
  K.~Koscher, A.~Czeskis, F.~Roesner, T.~Kohno, {\em et~al.}, ``{Comprehensive
  experimental analyses of automotive attack surfaces},'' in {\em USENIX
  Security Symposium}, vol.~4, (San Francisco), pp.~447--462, 2011.

\bibitem{koscher2010experimental}
K.~Koscher, A.~Czeskis, F.~Roesner, S.~Patel, T.~Kohno, S.~Checkoway, D.~McCoy,
  B.~Kantor, D.~Anderson, H.~Shacham, {\em et~al.}, ``Experimental security
  analysis of a modern automobile,'' in {\em IEEE Symposium on Security and
  Privacy}, pp.~447--462, IEEE, 2010.

\bibitem{yadav2016security}
A.~Yadav, G.~Bose, R.~Bhange, K.~Kapoor, N.~Iyengar, and R.~D. Caytiles,
  ``Security, vulnerability and protection of vehicular on-board diagnostics,''
  {\em International Journal of Security and Its Applications}, vol.~10, no.~4,
  pp.~405--422, 2016.

\bibitem{petit2015remote}
J.~Petit, B.~Stottelaar, M.~Feiri, and F.~Kargl, ``{Remote attacks on automated
  vehicles sensors: Experiments on camera and lidar},'' in {\em Black Hat
  Europe}, vol.~11, 2015.

\bibitem{yan2016can}
C.~Yan, W.~Xu, and J.~Liu, ``Can you trust autonomous vehicles: Contactless
  attacks against sensors of self-driving vehicle,'' in {\em DEF CON 24}, 2016.

\bibitem{cao2019adversarial}
Y.~Cao, C.~Xiao, B.~Cyr, Y.~Zhou, W.~Park, S.~Rampazzi, Q.~A. Chen, K.~Fu, and
  Z.~M. Mao, ``Adversarial sensor attack on lidar-based perception in
  autonomous driving,'' in {\em ACM SIGSAC Conference on Computer and
  Communications Security}, pp.~2267--2281, ACM, 2019.

\bibitem{Tesla3Spoof}
{Regulus Cyber LTD}, ``Tesla model 3 spoofed off the highway – regulus
  navigation system hack causes car to turn on its own.'' [Online]. Available:
  \url{https://www.regulus.com/blog/tesla-model-3-spoofed-off-the-highway-regulus-researches-hack-navigation-system-causing-car-to-steer-off-road/}.
  Accessed: July 11, 2020.

\bibitem{yang2016gnss}
Y.~Yang and J.~Xu, ``{GNSS receiver autonomous integrity monitoring (RAIM)
  algorithm based on robust estimation},'' {\em Geodesy and geodynamics},
  vol.~7, no.~2, pp.~117--123, 2016.

\bibitem{o2013real}
B.~W. O'Hanlon, M.~L. Psiaki, J.~A. Bhatti, D.~P. Shepard, and T.~E. Humphreys,
  ``{Real-Time GPS Spoofing Detection via Correlation of Encrypted Signals},''
  {\em Navigation}, vol.~60, no.~4, pp.~267--278, 2013.

\bibitem{wang2015pseudorandom}
Q.~Wang, Y.~Zhang, Y.~Xu, L.~Hao, Z.~Zhang, T.~Qiao, and Y.~Zhao,
  ``Pseudorandom modulation quantum secured lidar,'' {\em Optik}, vol.~126,
  no.~22, pp.~3344--3348, 2015.

\bibitem{nouri2017target}
M.~Nouri, M.~Mivehchy, and M.~F. Sabahi, ``Target recognition based on phase
  noise of received laser signal in lidar jammer,'' {\em Chinese Optics
  Letters}, vol.~15, no.~10, p.~100302, 2017.

\bibitem{sitawarin2018darts}
C.~Sitawarin, A.~N. Bhagoji, A.~Mosenia, M.~Chiang, and P.~Mittal, ``{DARTS:
  Deceiving Autonomous Cars with Toxic Signs},'' {\em arXiv preprint
  arXiv:1802.06430}, 2018.

\bibitem{man2019poster}
Y.~Man, M.~Li, and R.~Gerdes, ``{Poster: Perceived Adversarial Examples},'' in
  {\em IEEE Symposium on Security and Privacy}, 2019.

\bibitem{mahmud2005secure}
S.~M. Mahmud, S.~Shanker, and I.~Hossain, ``Secure software upload in an
  intelligent vehicle via wireless communication links,'' in {\em IEEE
  Proceedings. Intelligent Vehicles Symposium}, pp.~588--593, IEEE, 2005.

\bibitem{kalmeshwar2017development}
M.~Kalmeshwar and K.~N. Prasad, ``{Development of On-Board Diagnostics for Car
  and it's Integration with Android Mobile},'' in {\em 2nd International
  Conference on Computational Systems and Information Technology for
  Sustainable Solution (CSITSS)}, pp.~1--6, IEEE, 2017.

\bibitem{lin2005development}
C.~E. Lin, C.-C. Li, S.-H. Yang, S.-h. Lin, and C.-y. Lin, ``Development of
  on-line diagnostics and real time early warning system for vehicles,'' in
  {\em Sensors for Industry Conference}, pp.~45--51, IEEE, 2005.

\bibitem{kassakian1996automotive}
J.~G. Kassakian, H.-C. Wolf, J.~M. Miller, and C.~J. Hurton, ``Automotive
  electrical systems circa 2005,'' {\em IEEE spectrum}, vol.~33, no.~8,
  pp.~22--27, 1996.

\bibitem{singh2009tire}
S.~Singh, K.~Kingsley, and C.~L. Chen, ``Tire pressure maintenance--a
  statistical investigation,'' tech. rep., National Highway Traffic Safety
  Administration, 2009.

\bibitem{regulation2009661}
{European Parliament, Council of the European Union}, ``{Regulation (EC) No
  661/2009 of the European Parliament and of the Council of 13 July 2009
  concerning type-approval requirements for the general safety of motor
  vehicles, their trailers and systems, components and separate technical units
  intended therefor}.'' {Official Journal of the European Union}, 2009.

\bibitem{jitpakdee2008neural}
R.~Jitpakdee and T.~Maneewarn, ``Neural networks terrain classification using
  inertial measurement unit for an autonomous vehicle,'' in {\em SICE Annual
  Conference}, pp.~554--558, IEEE, 2008.

\bibitem{hpl2002introduction}
S.~Corrigan, ``{Introduction to the controller area network (CAN)},'' tech.
  rep., {Texas Instruments}, 2002.

\bibitem{wandinger2005introduction}
U.~Wandinger, ``{Introduction to LiDAR},'' in {\em {LiDAR: Range-Resolved
  Optical Remote Sensing of the Atmosphere}} (C.~Weitkamp, ed.), pp.~1--18,
  Springer, 2005.

\bibitem{hecht2018lidar}
J.~Hecht, ``Lidar for self-driving cars,'' {\em Optics and Photonics News},
  vol.~29, no.~1, pp.~26--33, 2018.

\bibitem{hulshof2013autonomous}
W.~Hulshof, I.~Knight, A.~Edwards, M.~Avery, and C.~Grover, ``Autonomous
  emergency braking test results,'' in {\em Proceedings of the 23rd
  International Technical Conference on the Enhanced Safety of Vehicles (ESV)},
  pp.~1--13, 2013.

\bibitem{uselmann2004sonic}
D.~J. Uselmann and L.~M. Uselmann, ``Sonic blind spot monitoring system,''
  Apr.~27 2004.
\newblock US Patent 6,727,808.

\bibitem{ishida2004development}
S.~Ishida and J.~E. Gayko, ``Development, evaluation and introduction of a lane
  keeping assistance system,'' in {\em IEEE Intelligent Vehicles Symposium},
  pp.~943--944, IEEE, 2004.

\bibitem{reed2003vehicle}
J.~C. Reed, ``Vehicle back-up and parking aid radar system,'' June~24 2003.
\newblock US Patent 6,583,753.

\bibitem{chi1992automatic}
C.~Chi, ``Automatic safety driving distance control device for a vehicle,''
  Nov.~24 1992.
\newblock US Patent 5,165,497.

\bibitem{breuer2007real}
J.~J. Breuer, A.~Faulhaber, P.~Frank, and S.~Gleissner, ``Real world safety
  benefits of brake assistance systems,'' in {\em 20th International Technical
  Conference on the Enhanced Safety of Vehicles (ESV)}, 2007.

\bibitem{team2014global}
{William J. Hughes Technical Center}, ``{Global positioning system (GPS)
  standard positioning service (SPS) performance analysis report},'' tech.
  rep., Federal Aviation Administration, 2014.

\bibitem{twitchell2001gps}
R.~W. Twitchell and A.~Taylor, ``{GPS location for mobile phones using the
  internet},'' Apr.~24 2001.
\newblock US Patent 6,222,483.

\bibitem{fairfield2011traffic}
N.~Fairfield and C.~Urmson, ``Traffic light mapping and detection,'' in {\em
  IEEE International Conference on Robotics and Automation}, pp.~5421--5426,
  IEEE, 2011.

\bibitem{omachi2009traffic}
M.~Omachi and S.~Omachi, ``Traffic light detection with color and edge
  information,'' in {\em 2nd IEEE International Conference on Computer Science
  and Information Technology}, pp.~284--287, IEEE, 2009.

\bibitem{levinson2011traffic}
J.~Levinson, J.~Askeland, J.~Dolson, and S.~Thrun, ``Traffic light mapping,
  localization, and state detection for autonomous vehicles,'' in {\em IEEE
  International Conference on Robotics and Automation}, pp.~5784--5791, IEEE,
  2011.

\bibitem{sun2013robust}
T.~Sun, S.~Tang, J.~Wang, and W.~Zhang, ``A robust lane detection method for
  autonomous car-like robot,'' in {\em Fourth International Conference on
  Intelligent Control and Information Processing (ICICIP)}, pp.~373--378, IEEE,
  2013.

\bibitem{hillel2014recent}
A.~B. Hillel, R.~Lerner, D.~Levi, and G.~Raz, ``Recent progress in road and
  lane detection: a survey,'' {\em Machine vision and applications}, vol.~25,
  no.~3, pp.~727--745, 2014.

\bibitem{wang2019pseudo}
Y.~Wang, W.-L. Chao, D.~Garg, B.~Hariharan, M.~Campbell, and K.~Q. Weinberger,
  ``Pseudo-lidar from visual depth estimation: Bridging the gap in 3d object
  detection for autonomous driving,'' in {\em Proceedings of the IEEE
  Conference on Computer Vision and Pattern Recognition}, pp.~8445--8453, 2019.

\bibitem{chang2015estimated}
J.~Chang, G.~Hatcher, D.~Hicks, J.~Schneeberger, B.~Staples, S.~Sundarajan,
  M.~Vasudevan, P.~Wang, K.~Wunderlich, {\em et~al.}, ``{Estimated benefits of
  connected vehicle applications: dynamic mobility applications, AERIS, V2I
  safety, and road weather management applications},'' tech. rep., United
  States Department of Transportation, 2015.

\bibitem{watfa2010advances}
M.~Watfa, {\em Advances in Vehicular Ad-Hoc Networks: Developments and
  Challenges: Developments and Challenges}.
\newblock IGI Global, 2010.

\bibitem{abboud2016interworking}
K.~Abboud, H.~A. Omar, and W.~Zhuang, ``{Interworking of DSRC and cellular
  network technologies for V2X communications: A survey},'' {\em IEEE
  transactions on vehicular technology}, vol.~65, no.~12, pp.~9457--9470, 2016.

\bibitem{baccelli2010ipv6}
E.~Baccelli, T.~Clausen, and R.~Wakikawa, ``{IPv6 operation for WAVE—Wireless
  access in vehicular environments},'' in {\em IEEE Vehicular Networking
  Conference}, pp.~160--165, IEEE, 2010.

\bibitem{kenney2011dedicated}
J.~B. Kenney, ``{Dedicated short-range communications (DSRC) standards in the
  United States},'' {\em Proceedings of the IEEE}, vol.~99, no.~7,
  pp.~1162--1182, 2011.

\bibitem{cho2016fingerprinting}
K.~T. Cho and K.~G. Shin, ``Fingerprinting electronic control units for vehicle
  intrusion detection,'' in {\em {25th USENIX Security Symposium}},
  pp.~911--927, 2016.

\bibitem{lin2012cyber}
C.~W. Lin and A.~Sangiovanni-Vincentelli, ``Cyber-security for the controller
  area network (can) communication protocol,'' in {\em International Conference
  on Cyber Security}, pp.~1--7, IEEE, 2012.

\bibitem{nilsson2008framework}
D.~K. Nilsson, L.~Sun, and T.~Nakajima, ``{A framework for self-verification of
  firmware updates over the air in vehicle ECUs},'' in {\em IEEE Globecom
  Workshops}, pp.~1--5, IEEE, 2008.

\bibitem{van2011canauth}
A.~Van~Herrewege, D.~Singelee, and I.~Verbauwhede, ``{CANAuth-a simple,
  backward compatible broadcast authentication protocol for CAN bus},'' in {\em
  ECRYPT Workshop on Lightweight Cryptography}, 2011.

\bibitem{halabi2018lightweight}
J.~Halabi and H.~Artail, ``A lightweight synchronous cryptographic hash chain
  solution to securing the vehicle can bus,'' in {\em IEEE International
  Multidisciplinary Conference on Engineering Technology (IMCET)}, pp.~1--6,
  IEEE, 2018.

\bibitem{uhlir2017practial}
D.~Uhlir, P.~Sedlacek, and J.~Hosek, ``{Practial overview of commercial
  connected cars systems in Europe},'' in {\em 9th International Congress on
  Ultra Modern Telecommunications and Control Systems and Workshops (ICUMT)},
  pp.~436--444, IEEE, 2017.

\bibitem{zamfir2019automotive}
S.~Zamfir and R.~Drosescu, ``Automotive black box and development platform used
  for traffic risks evaluation and mitigation,'' in {\em SIAR International
  Congress of Automotive and Transport Engineering: Science and Management of
  Automotive and Transportation Engineering}, pp.~426--438, Springer, 2019.

\bibitem{marstorp2017security}
G.~Marstorp and H.~Lindstr{\"o}m, ``{Security Testing of an OBD-II Connected
  IoT Device}.''
  \url{http://autosec.se/wp-content/uploads/2018/05/Marstorp-Lindstrom-Security-Testing-of-an-OBD-II-Connected-IoT-Device.pdf},
  2017.

\bibitem{christensen2019ethical}
L.~Christensen and D.~Dannberg, ``{Ethical hacking of IoT devices: OBD-II
  dongles}.''
  \url{http://www.diva-portal.org/smash/get/diva2:1333813/FULLTEXT01.pdf},
  2019.

\bibitem{ishtiaq2010security}
R.~M. Ishtiaq~Roufa, H.~Mustafaa, S.~O. Travis~Taylora, W.~Xua, M.~Gruteserb,
  W.~Trappeb, and I.~Seskarb, ``Security and privacy vulnerabilities of in-car
  wireless networks: A tire pressure monitoring system case study,'' in {\em
  19th USENIX Security Symposium}, pp.~11--13, 2010.

\bibitem{miller2013adventures}
C.~Miller and C.~Valasek, ``Adventures in automotive networks and control
  units,'' in {\em Def Con 21 Archive}, pp.~260--264, 2013.

\bibitem{woo2014practical}
S.~Woo, H.~J. Jo, and D.~H. Lee, ``A practical wireless attack on the connected
  car and security protocol for in-vehicle can,'' {\em IEEE Transactions on
  intelligent transportation systems}, vol.~16, no.~2, pp.~993--1006, 2014.

\bibitem{fowler2017towards}
D.~S. Fowler, M.~Cheah, S.~A. Shaikh, and J.~Bryans, ``Towards a testbed for
  automotive cybersecurity,'' in {\em IEEE International Conference on Software
  Testing, Verification and Validation (ICST)}, pp.~540--541, IEEE, 2017.

\bibitem{matsumoto2012method}
T.~Matsumoto, M.~Hata, M.~Tanabe, K.~Yoshioka, and K.~Oishi, ``A method of
  preventing unauthorized data transmission in controller area network,'' in
  {\em IEEE 75th Vehicular Technology Conference (VTC Spring)}, pp.~1--5, IEEE,
  2012.

\bibitem{hoppe2008security}
T.~Hoppe, S.~Kiltz, and J.~Dittmann, ``{Security threats to automotive CAN
  networks--practical examples and selected short-term countermeasures},'' in
  {\em International Conference on Computer Safety, Reliability, and Security},
  pp.~235--248, Springer, 2008.

\bibitem{wolf2006secure}
M.~Wolf, A.~Weimerskirch, and C.~Paar, ``Secure in-vehicle communication,'' in
  {\em Embedded Security in Cars}, pp.~95--109, Springer, 2006.

\bibitem{nilsson2008efficient}
D.~K. Nilsson, U.~E. Larson, and E.~Jonsson, ``Efficient in-vehicle delayed
  data authentication based on compound message authentication codes,'' in {\em
  IEEE 68th Vehicular Technology Conference}, pp.~1--5, IEEE, 2008.

\bibitem{muter2011entropy}
M.~M{\"u}ter and N.~Asaj, ``Entropy-based anomaly detection for in-vehicle
  networks,'' in {\em IEEE Intelligent Vehicles Symposium (IV)},
  pp.~1110--1115, IEEE, 2011.

\bibitem{shin2017fingerprinting}
K.~G. Shin and K.~T. Cho, ``Fingerprinting electronic control units for vehicle
  intrusion detection,'' Oct.~5 2017.
\newblock US Patent App. 15/472,861.

\bibitem{gmiden2016intrusion}
M.~Gmiden, M.~H. Gmiden, and H.~Trabelsi, ``{An intrusion detection method for
  securing in-vehicle CAN bus},'' in {\em 17th International Conference on
  Sciences and Techniques of Automatic Control and Computer Engineering (STA)},
  pp.~176--180, IEEE, 2016.

\bibitem{tyree2018exploiting}
Z.~Tyree, R.~A. Bridges, F.~L. Combs, and M.~R. Moore, ``{Exploiting the shape
  of CAN data for in-vehicle intrusion detection},'' in {\em IEEE 88th
  Vehicular Technology Conference (VTC-Fall)}, pp.~1--5, IEEE, 2018.

\bibitem{siddiqui2017secure}
A.~S. Siddiqui, Y.~Gui, J.~Plusquellic, and F.~Saqib, ``Secure communication
  over canbus,'' in {\em 2017 IEEE 60th International Midwest Symposium on
  Circuits and Systems (MWSCAS)}, pp.~1264--1267, IEEE, 2017.

\bibitem{seshadri2006scuba}
A.~Seshadri, M.~Luk, A.~Perrig, L.~van Doorn, and P.~Khosla, ``{SCUBA: Secure
  code update by attestation in sensor networks},'' in {\em Proceedings of the
  5th ACM workshop on Wireless security}, pp.~85--94, ACM, 2006.

\bibitem{prathap2013penetration}
V.~Prathap and A.~RACHUMALLU, ``{Penetration Testing of Vehicle ECUs},''
  Master's thesis, Chalmers University of Technology, 2013.

\bibitem{nilsson2008creating}
D.~K. Nilsson, U.~E. Larson, and E.~Jonsson, ``Creating a secure infrastructure
  for wireless diagnostics and software updates in vehicles,'' in {\em
  International Conference on Computer Safety, Reliability, and Security},
  pp.~207--220, Springer, 2008.

\bibitem{nilsson2008secure}
D.~K. Nilsson and U.~E. Larson, ``Secure firmware updates over the air in
  intelligent vehicles,'' in {\em ICC Workshops - IEEE International Conference
  on Communications Workshops}, pp.~380--384, IEEE, 2008.

\bibitem{mulliner2012read}
C.~Mulliner and B.~Mich{\'e}le, ``{Read It Twice! A Mass-Storage-Based TOCTTOU
  Attack},'' in {\em 6th USENIX Workshop on Offensive Technologies (WOOT 12)},
  pp.~105--112, 2012.

\bibitem{shin2017illusion}
H.~Shin, D.~Kim, Y.~Kwon, and Y.~Kim, ``Illusion and dazzle: Adversarial
  optical channel exploits against lidars for automotive applications,'' in
  {\em International Conference on Cryptographic Hardware and Embedded
  Systems}, pp.~445--467, Springer, 2017.

\bibitem{pace2009detecting}
P.~E. Pace, {\em Detecting and classifying low probability of intercept radar}.
\newblock Artech House, 2009.

\bibitem{davidson2016controlling}
D.~Davidson, H.~Wu, R.~Jellinek, V.~Singh, and T.~Ristenpart, ``{Controlling
  UAVs with sensor input spoofing attacks},'' in {\em {10th USENIX Workshop on
  Offensive Technologies (WOOT 16)}}, 2016.

\bibitem{matsumura2018secure}
R.~Matsumura, T.~Sugawara, and K.~Sakiyama, ``{A secure LiDAR with AES-based
  side-channel fingerprinting},'' in {\em Sixth International Symposium on
  Computing and Networking Workshops (CANDARW)}, pp.~479--482, IEEE, 2018.

\bibitem{foxnews}
{Fox News}, ``{Zapped: Driver fined thousands for using laser jammer}.''
  [Online]. Available:
  \url{https://www.foxnews.com/auto/zapped-driver-fined-thousands-for-using-laser-jammer}.
  Accessed: July 11, 2020.

\bibitem{stottelaar2015practical}
B.~G. Stottelaar, ``Practical cyber-attacks on autonomous vehicles,'' Master's
  thesis, University of Twente, 2015.

\bibitem{roome1990digital}
S.~Roome, ``Digital radio frequency memory,'' {\em Electronics \& communication
  engineering journal}, vol.~2, no.~4, pp.~147--153, 1990.

\bibitem{chauhan2014platform}
R.~Chauhan, ``A platform for false data injection in frequency modulated
  continuous wave radar,'' Master's thesis, Utah State University, 2014.

\bibitem{shoukry2015pycra}
Y.~Shoukry, P.~Martin, Y.~Yona, S.~Diggavi, and M.~Srivastava, ``Pycra:
  Physical challenge-response authentication for active sensors under spoofing
  attacks,'' in {\em Proceedings of the 22nd ACM SIGSAC Conference on Computer
  and Communications Security}, pp.~1004--1015, ACM, 2015.

\bibitem{kapoor2018detecting}
P.~Kapoor, A.~Vora, and K.-D. Kang, ``Detecting and mitigating spoofing attack
  against an automotive radar,'' in {\em IEEE 88th Vehicular Technology
  Conference (VTC-Fall)}, pp.~1--6, IEEE, 2018.

\bibitem{dutta2017estimation}
R.~G. Dutta, X.~Guo, T.~Zhang, K.~Kwiat, C.~Kamhoua, L.~Njilla, and Y.~Jin,
  ``Estimation of safe sensor measurements of autonomous system under attack,''
  in {\em Proceedings of the 54th Annual Design Automation Conference 2017},
  pp.~1--6, 2017.

\bibitem{buehler2014airborne}
W.~E. Buehler, R.~M. Whitson, and M.~J. Lewis, ``Airborne radar jamming
  system,'' Sept.~9 2014.
\newblock US Patent 8,830,112.

\bibitem{stott1994digital}
G.~Stott, ``Digital modulation for radar jamming,'' in {\em IEE Colloquium on
  Signal Processing in Electronic Warfare}, pp.~1--1, IET, 1994.

\bibitem{Lothes1990jammingbook}
R.~N. Lothes, M.~B. Szymanski, and R.~G. Wiley, {\em {Radar Vulnerability to
  Jamming}}.
\newblock Artech House, 1990.

\bibitem{greco2005combined}
M.~Greco, F.~Gini, and A.~Farina, ``{Combined effect of phase and RGPO delay
  quantization on jamming signal spectrum},'' in {\em IEEE International Radar
  Conference}, pp.~37--42, IEEE, 2005.

\bibitem{greco2008radar}
M.~Greco, F.~Gini, and A.~Farina, ``Radar detection and classification of
  jamming signals belonging to a cone class,'' {\em IEEE transactions on signal
  processing}, vol.~56, no.~5, pp.~1984--1993, 2008.

\bibitem{lu2010anti}
G.~Lu, D.~Zeng, and B.~Tang, ``{Anti-jamming filtering for DRFM repeat jammer
  based on stretch processing},'' in {\em 2nd International Conference on
  Signal Processing Systems}, vol.~1, pp.~V1--78, IEEE, 2010.

\bibitem{krasner2000method}
N.~F. Krasner, ``{Method and apparatus for adaptively processing GPS signals in
  a GPS receiver},'' Oct.~17 2000.
\newblock US Patent 6,133,873.

\bibitem{tippenhauer2011requirements}
N.~O. Tippenhauer, C.~P{\"o}pper, K.~B. Rasmussen, and S.~Capkun, ``{On the
  requirements for successful GPS spoofing attacks},'' in {\em 18th ACM
  conference on Computer and communications security}, pp.~75--86, ACM, 2011.

\bibitem{psiaki2014gnss}
M.~L. Psiaki, B.~W. O'hanlon, S.~P. Powell, J.~A. Bhatti, K.~D. Wesson, and
  T.~E. Humphreys, ``Gnss spoofing detection using two-antenna differential
  carrier phase,'' in {\em Radionavigation Laboratory Conference Proceedings},
  2014.

\bibitem{shepard2012evaluation}
D.~P. Shepard, T.~E. Humphreys, and A.~A. Fansler, ``{Evaluation of the
  vulnerability of phasor measurement units to GPS spoofing attacks},'' {\em
  International Journal of Critical Infrastructure Protection}, vol.~5,
  no.~3-4, pp.~146--153, 2012.

\bibitem{zeng2018all}
K.~C. Zeng, S.~Liu, Y.~Shu, D.~Wang, H.~Li, Y.~Dou, G.~Wang, and Y.~Yang,
  ``{All your GPS are belong to us: Towards stealthy manipulation of road
  navigation systems},'' in {\em {27th USENIX Security Symposium (USENIX
  Security 18)}}, pp.~1527--1544, 2018.

\bibitem{narain2019security}
S.~Narain, A.~Ranganathan, and G.~Noubir, ``{Security of GPS/INS based on-road
  location tracking systems},'' in {\em IEEE Symposium on Security and Privacy
  (SP)}, pp.~587--601, IEEE, 2019.

\bibitem{petovello2001development}
M.~Petovello, M.~Cannon, G.~Lachapelle, J.~Wang, C.~Wilson, O.~Salychev, and
  V.~Voronov, ``{Development and testing of a real-time GPS/INS reference
  system for autonomous automobile navigation},'' in {\em Proceedings of ION
  GPS}, vol.~1, pp.~2634--2641, 2001.

\bibitem{el2006kalman}
N.~El-Sheimy, E.~H. Shin, and X.~Niu, ``{Kalman filter face-off: Extended vs.
  unscented Kalman filters for integrated GPS and MEMS inertial},'' {\em Inside
  GNSS}, vol.~1, no.~2, pp.~48--54, 2006.

\bibitem{meng2019gps}
Q.~Meng, L.~T. Hsu, B.~Xu, X.~Luo, and A.~El-Mowafy, ``{A GPS Spoofing
  Generator Using an Open Sourced Vector Tracking-Based Receiver},'' {\em
  Sensors}, vol.~19, no.~18, p.~3993, 2019.

\bibitem{warner2003gps}
J.~S. Warner and R.~G. Johnston, ``{GPS spoofing countermeasures},'' {\em
  Homeland Security Journal}, vol.~25, no.~2, pp.~19--27, 2003.

\bibitem{brown1996receiver}
R.~G. Brown, ``Receiver autonomous integrity monitoring,'' {\em Global
  Positioning System: Theory and applications.}, vol.~2, pp.~143--165, 1996.

\bibitem{hewitson2006gnss}
S.~Hewitson and J.~Wang, ``{GNSS receiver autonomous integrity monitoring
  (RAIM) performance analysis},'' {\em GPS Solutions}, vol.~10, no.~3,
  pp.~155--170, 2006.

\bibitem{van1992raim}
K.~L. Van~Dyke, ``{RAIM availability for supplemental GPS navigation},'' {\em
  Navigation}, vol.~39, no.~4, pp.~429--443, 1992.

\bibitem{loh1994integrity}
R.~Loh and J.~Fernow, ``{Integrity monitoring requirements for FAA's GPS
  wide-area augmentation system (WAAS)},'' in {\em IEEE Position, Location and
  Navigation Symposium}, pp.~629--636, IEEE, 1994.

\bibitem{blanch2012advanced}
J.~Blanch, T.~Walter, P.~Enge, Y.~Lee, B.~Pervan, M.~Rippl, and A.~Spletter,
  ``{Advanced RAIM user algorithm description: integrity support message
  processing, fault detection, exclusion, and protection level calculation},''
  in {\em Proceedings of the 25th International Technical Meeting of The
  Satellite Division of the Institute of Navigation (ION GNSS 2012)},
  pp.~2828--2849, 2012.

\bibitem{choi2011advanced}
M.~Choi, J.~Blanch, T.~Walter, and P.~Enge, ``{Advanced RAIM demonstration
  using four months of ground data},'' {\em Proceedings of the 2011
  International Technical Meeting of The Institute of Navigation},
  pp.~279--284, January 2011.

\bibitem{qian2019impact}
M.~Qian, L.~Jianye, Z.~Qinghua, F.~Shaojun, and X.~Rui, ``{Impact of one
  satellite outage on ARAIM depleted constellation configurations},'' {\em
  Chinese Journal of Aeronautics}, vol.~32, no.~4, pp.~967--977, 2019.

\bibitem{meng2019improved}
Q.~Meng, J.~Liu, Q.~Zeng, S.~Feng, and R.~Xu, ``{Improved ARAIM fault modes
  determination scheme based on feedback structure with probability
  accumulation},'' {\em GPS Solutions}, vol.~23, no.~1, p.~16, 2019.

\bibitem{montgomery2011receiver}
P.~Y. Montgomery, ``{Receiver-autonomous spoofing detection: Experimental
  results of a multi-antenna receiver defense against a portable civil GPS
  spoofer},'' in {\em Radionavigation Laboratory Conference Proceedings}, 2011.

\bibitem{hu2009study}
H.~Hu and N.~Wei, ``{A study of GPS jamming and anti-jamming},'' in {\em 2nd
  international conference on power electronics and intelligent transportation
  system (PEITS)}, vol.~1, pp.~388--391, IEEE, 2009.

\bibitem{helfrick2014question}
A.~Helfrick, ``{Question: Alternate position, navigation timing, APNT? Answer:
  ELORAN},'' in {\em IEEE/AIAA 33rd Digital Avionics Systems Conference
  (DASC)}, pp.~3C3--1--3C3--9, IEEE, 2014.

\bibitem{coffed2014threat}
J.~Coffed, ``{The threat of GPS jamming: The risk to an information utility},''
  tech. rep., Harris Corporation, 2014.

\bibitem{hunkeler2012effectiveness}
U.~Hunkeler, J.~Colli-Vignarelli, and C.~Dehollain, ``{Effectiveness of
  GPS-jamming and counter-measures},'' in {\em International Conference on
  Localization and GNSS}, pp.~1--4, IEEE, 2012.

\bibitem{zhang2012anti}
Y.~D. Zhang and M.~G. Amin, ``{Anti-jamming GPS receiver with reduced phase
  distortions},'' {\em IEEE Signal Processing Letters}, vol.~19, no.~10,
  pp.~635--638, 2012.

\bibitem{sun2005self}
W.~Sun and M.~G. Amin, ``{A self-coherence anti-jamming GPS receiver},'' {\em
  IEEE Transactions on Signal Processing}, vol.~53, no.~10, pp.~3910--3915,
  2005.

\bibitem{pattinson2017standardisation}
M.~Pattinson, M.~Dumville, Y.~Ying, D.~Fryganiotis, M.~Z.~H. Bhuiyan,
  S.~Thombre, H.~Kuusniemi, A.~Waern, P.~Eliardsson, S.~Hill, V.~Manikundalam,
  S.~Lee, and J.~Gonzalez, ``{Standardization of GNSS Threat reporting and
  Receiver testing through International Knowledge Exchange, Experimentation
  and Exploitation [STRIKE3] - Draft Standards for Receiver Testing},'' {\em
  European Journal of Navigation}, vol.~15, December 2017.

\bibitem{mukhopadhyay2007augmentation}
M.~Mukhopadhyay, B.~K. Sarkar, and A.~Chakraborty, ``{Augmentation of anti-jam
  GPS system using smart antenna with a simple DOA estimation algorithm},''
  {\em Progress In Electromagnetics Research}, vol.~67, pp.~231--249, 2007.

\bibitem{purwar2016gps}
A.~Purwar, D.~Joshi, and V.~K. Chaubey, ``{GPS signal jamming and anti-jamming
  strategy—A theoretical analysis},'' in {\em IEEE Annual India Conference
  (INDICON)}, pp.~1--6, IEEE, 2016.

\bibitem{dh2020autonomous}
D.~H. Sharath~Yadav and A.~Ansari, ``{Autonomous Vehicles Camera Blinding
  Attack Detection Using Sequence Modelling and Predictive Analytics},'' tech.
  rep., SAE International. Available at
  \url{https://doi.org/10.4271/2020-01-0719}., 2020.

\bibitem{brown2017adversarial}
T.~B. Brown, D.~Man{\'e}, A.~Roy, M.~Abadi, and J.~Gilmer, ``Adversarial
  patch,'' {\em arXivpreprint arXiv:1712.09665}, 2018.

\bibitem{lu2017no}
J.~Lu, H.~Sibai, E.~Fabry, and D.~Forsyth, ``No need to worry about adversarial
  examples in object detection in autonomous vehicles,'' {\em arXiv preprint
  arXiv:1707.03501}, 2017.

\bibitem{eykholt2018robust}
K.~Eykholt, I.~Evtimov, E.~Fernandes, B.~Li, A.~Rahmati, C.~Xiao, A.~Prakash,
  T.~Kohno, and D.~Song, ``Robust physical-world attacks on deep learning
  visual classification,'' in {\em Proceedings of the IEEE Conference on
  Computer Vision and Pattern Recognition}, pp.~1625--1634, 2018.

\bibitem{kelarestaghi2019intelligent}
K.~B. Kelarestaghi, M.~Foruhandeh, K.~Heaslip, and R.~Gerdes, ``Intelligent
  transportation system security: Impact-oriented risk assessment of in-vehicle
  networks,'' {\em IEEE Intelligent Transportation Systems Magazine}, pp.~1--1,
  2019.

\bibitem{moosavi2017universal}
S.~M. Moosavi-Dezfooli, A.~Fawzi, O.~Fawzi, and P.~Frossard, ``Universal
  adversarial perturbations,'' in {\em Proceedings of the IEEE conference on
  computer vision and pattern recognition}, pp.~1765--1773, 2017.

\bibitem{goodfellow2014explaining}
I.~J. Goodfellow, J.~Shlens, and C.~Szegedy, ``Explaining and harnessing
  adversarial examples,'' {\em arXiv preprint arXiv:1412.6572}, 2014.

\bibitem{kos2018adversarial}
J.~Kos, I.~Fischer, and D.~Song, ``Adversarial examples for generative
  models,'' in {\em IEEE Security and Privacy Workshops (SPW)}, pp.~36--42,
  IEEE, 2018.

\bibitem{art2018}
M.~I. Nicolae, M.~Sinn, M.~N. Tran, B.~Buesser, A.~Rawat, M.~Wistuba,
  V.~Zantedeschi, N.~Baracaldo, B.~Chen, H.~Ludwig, I.~Molloy, and B.~Edwards,
  ``Adversarial robustness toolbox v1.0.1,'' {\em Computing Research Repository
  (CoRR)}, vol.~1807.01069, 2018.

\bibitem{zantedeschi2017efficient}
V.~Zantedeschi, M.~I. Nicolae, and A.~Rawat, ``Efficient defenses against
  adversarial attacks,'' in {\em Proceedings of the 10th ACM Workshop on
  Artificial Intelligence and Security}, pp.~39--49, ACM, 2017.

\bibitem{xu2017feature}
W.~Xu, D.~Evans, and Y.~Qi, ``Feature squeezing: Detecting adversarial examples
  in deep neural networks,'' {\em arXiv preprint arXiv:1704.01155}, 2017.

\bibitem{dziugaite2016study}
G.~K. Dziugaite, Z.~Ghahramani, and D.~M. Roy, ``A study of the effect of jpg
  compression on adversarial images,'' {\em arXiv preprint arXiv:1608.00853},
  2016.

\bibitem{guo2017countering}
C.~Guo, M.~Rana, M.~Cisse, and L.~Van Der~Maaten, ``Countering adversarial
  images using input transformations,'' {\em arXiv preprint arXiv:1711.00117},
  2017.

\bibitem{szegedy2013intriguing}
C.~Szegedy, W.~Zaremba, I.~Sutskever, J.~Bruna, D.~Erhan, I.~Goodfellow, and
  R.~Fergus, ``Intriguing properties of neural networks,'' {\em arXiv preprint
  arXiv:1312.6199}, 2013.

\bibitem{miyato2015distributional}
T.~Miyato, S.~Maeda, M.~Koyama, K.~Nakae, and S.~Ishii, ``Distributional
  smoothing with virtual adversarial training,'' {\em arXiv preprint
  arXiv:1507.00677}, 2015.

\bibitem{jiang2020poisoning}
W.~Jiang, H.~Li, S.~Liu, X.~Luo, and R.~Lu, ``Poisoning and evasion attacks
  against deep learning algorithms in autonomous vehicles,'' {\em IEEE
  transactions on vehicular technology}, vol.~69, no.~4, pp.~4439--4449, 2020.

\bibitem{buckman2018thermometer}
J.~Buckman, A.~Roy, C.~Raffel, and I.~Goodfellow, ``Thermometer encoding: One
  hot way to resist adversarial examples,'' in {\em 6th International
  Conference on Learning Representations, {ICLR}}, 2018.

\bibitem{chim2009security}
T.~W. Chim, S.~M. Yiu, L.~C.~K. Hui, and V.~O.~K. Li, ``{Security and privacy
  issues for inter-vehicle communications in VANETs},'' in {\em 6th IEEE Annual
  Communications Society Conference on Sensor, Mesh and Ad Hoc Communications
  and Networks Workshops}, pp.~1--3, IEEE, 2009.

\bibitem{rawat2012vanet}
A.~Rawat, S.~Sharma, and R.~Sushil, ``{VANET: Security attacks and its possible
  solutions},'' {\em Journal of Information and Operations Management}, vol.~3,
  no.~1, p.~301, 2012.

\bibitem{grover2011sybil}
J.~Grover, M.~S. Gaur, V.~Laxmi, and N.~K. Prajapati, ``A sybil attack
  detection approach using neighboring vehicles in vanet,'' in {\em Proceedings
  of the 4th international conference on Security of information and networks},
  pp.~151--158, ACM, 2011.

\bibitem{whyte2013security}
W.~Whyte, A.~Weimerskirch, V.~Kumar, and T.~Hehn, ``A security credential
  management system for v2v communications,'' in {\em IEEE Vehicular Networking
  Conference}, pp.~1--8, IEEE, 2013.

\bibitem{alimohammadi2015sybil}
M.~Alimohammadi and A.~A. Pouyan, ``{Sybil attack detection using a low cost
  short group signature in VANET},'' in {\em 12th International Iranian Society
  of Cryptology Conference on Information Security and Cryptology (ISCISC)},
  pp.~23--28, IEEE, 2015.

\bibitem{hubaux2004security}
J.~P. Hubaux, S.~Capkun, and J.~Luo, ``The security and privacy of smart
  vehicles,'' {\em IEEE Security \& Privacy}, no.~3, pp.~49--55, 2004.

\bibitem{zhang2008efficient}
C.~Zhang, R.~Lu, X.~Lin, P.~H. Ho, and X.~Shen, ``An efficient identity-based
  batch verification scheme for vehicular sensor networks,'' in {\em 27th
  Conference on Computer Communications}, pp.~246--250, IEEE, 2008.

\bibitem{zhao2020efficient}
M.~Zhao, D.~Qin, R.~Guo, and G.~Xu, ``Efficient protection mechanism based on
  self-adaptive decision for communication networks of autonomous vehicles,''
  {\em Mobile Information Systems}, 2020.

\bibitem{pathre2013identification}
A.~Pathre, ``{Identification of malicious vehicle in VANET environment from
  DDoS attack},'' {\em Journal of Global Research in Computer Science}, vol.~4,
  no.~6, pp.~30--34, 2013.

\bibitem{pathre2013novel}
A.~Pathre, C.~Agrawal, and A.~Jain, ``{A novel defense scheme against DDOS
  attack in VANET},'' in {\em Tenth International Conference on Wireless and
  Optical Communications Networks (WOCN)}, pp.~1--5, IEEE, 2013.

\bibitem{douligeris2004ddos}
C.~Douligeris and A.~Mitrokotsa, ``{DDoS attacks and defense mechanisms:
  classification and state-of-the-art},'' {\em Computer Networks}, vol.~44,
  no.~5, pp.~643--666, 2004.

\bibitem{hasbullah2010denial}
H.~Hasbullah, I.~A. Soomro, {\em et~al.}, ``{Denial of service (DOS) attack and
  its possible solutions in VANET},'' {\em International Journal of Electronics
  and Communication Engineering}, vol.~4, no.~5, pp.~813--817, 2010.

\bibitem{Ekedebe2015simulation}
N.~Ekedebe, W.~Yu, H.~Song, and C.~Lu, ``{On a simulation study of cyber
  attacks on vehicle-to-infrastructure communication (V2I) in Intelligent
  Transportation System (ITS)},'' in {\em Mobile Multimedia/Image Processing,
  Security, and Applications} (S.~S. Agaian, S.~A. Jassim, and E.~Y. Du, eds.),
  vol.~9497, pp.~96 -- 107, International Society for Optics and Photonics,
  SPIE, 2015.

\bibitem{blum2006fast}
J.~J. Blum and A.~Eskandarian, ``Fast, robust message forwarding for
  inter-vehicle communication networks,'' in {\em IEEE Intelligent
  Transportation Systems Conference}, pp.~1418--1423, IEEE, 2006.

\bibitem{leinmuller2006position}
T.~Leinmuller, E.~Schoch, and F.~Kargl, ``Position verification approaches for
  vehicular ad hoc networks,'' {\em IEEE Wireless Communications}, vol.~13,
  no.~5, pp.~16--21, 2006.

\bibitem{singh2018ml}
P.~K. Singh, S.~K. Jha, S.~K. Nandi, and S.~Nandi, ``{ML-Based Approach to
  Detect DDoS Attack in V2I Communication Under SDN Architecture},'' in {\em
  TENCON IEEE Region 10 Conference}, pp.~0144--0149, IEEE, 2018.

\bibitem{plossl2006towards}
K.~Plossl, T.~Nowey, and C.~Mletzko, ``Towards a security architecture for
  vehicular ad hoc networks,'' in {\em First International Conference on
  Availability, Reliability and Security (ARES'06)}, pp.~8--pp, IEEE, 2006.

\bibitem{techxplore}
{University of Warwick}, ``Cyber security of connected autonomous vehicles
  trialled.'' [Online]. Available:
  \url{https://techxplore.com/news/2019-09-cyber-autonomous-vehicles-trialled.html}.
  Accessed: July 11, 2020.

\bibitem{safetyfirst2019}
M.~Wood, P.~Robbel, M.~Maass, R.~Tebbens, M.~Meijs, M.~Harb, and P.~Schlicht,
  ``Safety first for automated driving.'' [Online]. Available:
  \url{https://newsroom.intel.com/wp-content/uploads/sites/11/2019/07/Intel-Safety-First-for-Automated-Driving.pdf},
  July 11, 2020.

\bibitem{selfdrivingcoalition}
{Self-Driving Coalition}, ``{Research Publications}.''
  \url{https://www.selfdrivingcoalition.org/resources/research}, {. Accessed
  June 2020}.

\bibitem{waymo2017report}
{Waymo LLC}, ``Waymo safety report - on the road to fully self-driving.''
  [Online]. Available:
  \url{https://storage.googleapis.com/sdc-prod/v1/safety-report/Safety\%20Report\%202018.pdf}.
  Accessed: July 11, 2020.

\bibitem{GuardKnox}
{Guardknox Cyber Technologies Ltd.}, ``{The Communication Lockdown™
  Methodology: FIGHTER JET CYBERSECURITY FOR CONNECTED VEHICLES}.''
  \url{https://learn.guardknox.com/communication-lockdown-whitepaper}.

\end{thebibliography}

\end{document}